\documentclass[lettersize, journal]{IEEEtran}
\usepackage[utf8]{inputenc}
\usepackage{cite}
\usepackage{amsmath,amssymb,amsfonts}
\usepackage{amsthm}
\usepackage{mathtools}
\allowdisplaybreaks
\usepackage[ruled,vlined]{algorithm2e}
\usepackage{graphicx}
\graphicspath{{figs/}{Uplink_ARS_Paper/figs/}}
\usepackage{setspace}
\usepackage{caption}
\usepackage{subcaption}
\usepackage{tikz}
\usepackage{tkz-euclide}
\usepackage[long]{optidef}
\usepackage{bm}
\usepackage{xcolor}
\usepackage[normalem]{ulem}
\usepackage{comment}
\usepackage{fancyhdr}
\usetikzlibrary{arrows.meta,positioning,fit,calc}
\usepackage[hidelinks]{hyperref}
\newif\ifshowchanges
\showchangesfalse

\newcommand{\R}{\mathbb{R}}
\newcommand{\Z}{\mathbb{Z}}
\newcommand{\E}{\mathbb{E}}
\newcommand{\C}{\mathcal{C}}
\newcommand{\ARS}{\mathrm{A\text{-}RS}}

\DeclareMathOperator{\diag}{diag}

\DeclareMathOperator{\rank}{rank}
\DeclareMathOperator{\tr}{tr}
\DeclareMathOperator{\spn}{span}
\DeclareMathOperator{\conv}{conv}
\newtheorem{theorem}{Theorem}
\newtheorem{proposition}{Proposition}
\newtheorem{lemma}{Lemma}
\newtheorem{corollary}{Corollary}

\theoremstyle{definition}
\newtheorem{definition}{Definition}

\fancypagestyle{firstpage}{%
  \fancyhf{} 
  \fancyhead[C]{%
    \begin{minipage}{\textwidth}
    \centering
    \scriptsize
    This work has been submitted to the IEEE for possible publication. 
    Copyright may be transferred without notice, after which this version may no longer be accessible.
    \end{minipage}%
  }%
}

\begin{document}

\title{Asynchronous Uplink Rate Splitting Multiple Access}
\author{\IEEEauthorblockN{Hossein Maleki, \IEEEmembership{Graduate Student Member, IEEE} and Hamid Jafarkhani, \IEEEmembership{Fellow, IEEE} \thanks{The authors are with the University of California, Irvine, CA, USA. This work was supported in part by the NSF Award CCF-2328075.}
}
}

\maketitle
\thispagestyle{firstpage}
\begin{abstract}
We study a two-user uplink system in which each user splits its message into layers and transmits them with deliberately chosen timing offsets, a scheme we call asynchronous rate splitting (A-RS). We show that timing offsets make the layers distinguishable only in the excess band of the pulse, where two spectral aliases overlap. For a root-raised-cosine pulse, the rate for every subset of layers has a scalar closed form that depends on the offsets through a single timing dispersion term. For fixed layer spectra and timing offsets, the achievable two-user rate region is a pentagon. Its two maximum-sum-rate corners are attained by successive interference cancellation that combines all matched-filter outputs at each decoding stage. Two layers per user, separated by half a symbol, are optimal for any delay between the users, and the spectrum of the user decoded last is a water-filling nested inside the water-filling of the total power. At high signal-to-noise ratio (SNR), the optimal spectrum has twice the density in the excess band. By using the excess bandwidth of the pulse, which synchronous transmission leaves unused, A-RS achieves rate pairs outside the regions of both synchronous and asynchronous non-orthogonal multiple access. We implement a practical system with low-density parity-check codes which gives the weaker user $0.97$ bits per symbol more than any single-stream scheme at the same SNR.
\end{abstract}


\section{Introduction}
Non-orthogonal multiple access (NOMA) lets several users share the same time and frequency resources and separates them at the receiver by successive interference cancellation (SIC)~\cite{saito2013noma,liu2017noma,10786246}. In the uplink, NOMA is a form of the Gaussian multiple-access channel (MAC), whose capacity region is the Cover--Wyner pentagon~\cite{wyner1974recent,cover1975advances}. SIC with time sharing attains every point of this pentagon~\cite{cover2006elements,tse2005fundamental}.

The conventional MAC design assumes that the symbol epochs of the users are aligned at the receiver. In practice, users are at different distances from the base station and their oscillators drift, so alignment requires timing compensation and closed-loop control, and residual offsets remain~\cite{nasir2016timing,abeywickrama2017ota}. Asynchronism can occur at two levels. If the codewords of the users do not start at the same time, the users cannot agree on when to switch between transmission strategies, so they lose the benefit of time sharing~\cite{hui1985asynchronous}, and over channels with memory even their total rate can decrease~\cite{verdu1989memory}. If the symbols of the users are also misaligned, their pulses overlap partially at the receiver, and the capacity region of the Gaussian MAC depends on how much the pulses overlap~\cite{verdu1989asynchronous}. Its boundary is attained when users shape their power over frequency by a multiuser water-filling~\cite{chengverdu1993waterfilling}, an extension of the single-user water-filling over colored noise~\cite{shannon1949communication} and intersymbol interference~\cite{hirt1988capacity}. Unlike the straight dominant face of the synchronous pentagon capacity region, this boundary is curved. Therefore, timing offset does not merely degrade the synchronous rate region but reshapes it and can improve it. Therefore, the choice of timing offset becomes a parameter of the communication channel.

This suggests that asynchrony need not be a nuisance. Intentional timing offsets have been used to improve detection in multi-antenna systems~\cite{shao2007vblast,das2011mimo} and to mitigate interference through sampling diversity~\cite{ganji2016interference}. Timing offsets have been exploited in downlink~\cite{cui2017asynchronous,ganji2019downlink} and uplink~\cite{haci2017asynchronous,zou2019analysis} NOMA. They have also been used in transmit beamforming~\cite{ganji2020beamforming}, differential decoding for multiple access~\cite{poorkasmaei2015asynchronous}, and cooperative transmission~\cite{6965635}. Similarly, intentional frequency offsets have been exploited in multi-carrier NOMA~\cite{ganji2019frequency}. For band-limited pulses, the capacity region of the asynchronous uplink MAC is larger than that of the synchronous MAC, as proved in~\cite{ganji2021rate}. Intentional delays have also been used to manage interference in cell-free networks~\cite{maleki2026cellfree} and to create parallel interference-free streams on a single antenna~\cite{maleki2026delay}.

Asynchronous transmission is also related to faster-than-Nyquist (FTN) signaling, which sends pulses faster than the Nyquist rate of their bandwidth~\cite{mazo1975faster,anderson2013faster}. For pulses with excess bandwidth, the capacity of FTN exceeds that of orthogonal signaling~\cite{rusek2009constrained}. In our proposed method, the two layers of the same user are separated by half a symbol form a pulse train at a rate $2/T$. Therefore, from a single user point of view, our scheme is an FTN transmission whose two streams are separated by SIC.

Rate splitting divides a user into virtual users. In the synchronous Gaussian MAC, it attains every point of the capacity region with single-user decoding and without time sharing~\cite{rimoldi1996,grant2001, jafarkhani2024modulation}. In the downlink, rate-splitting multiple access (RSMA) uses the same idea to bridge and generalize space-division multiple access and NOMA~\cite{mao2018rsma,clerckx2020rate,mao2022}. However, in the synchronous MAC, splitting does not enlarge the region, since all layers of a user occupy the same signal dimension. In this article, we show that the situation is different when splitting is combined with asynchrony. Layers of the same user with different timing offsets occupy different signal dimensions in the excess band of the pulse. Therefore, asynchronous rate splitting (A-RS) enlarges the region beyond what any single-stream scheme can achieve, synchronous or asynchronous. Our contributions are as follows.
\begin{itemize}
\item We express the layered channel in the frequency domain and show that the layers occupy only as many signal dimensions as there are spectral copies of the pulse at each frequency. A root-raised-cosine pulse has two copies in its excess band, so timing offsets can separate layers in that region.
\item Using the block Szeg\H{o} theorem~\cite{grenander1958toeplitz,gray2006toeplitz,gutierrez2008toeplitz}, we show that the layered channel separates over frequency, i.e., at each frequency, the layers see a memoryless Gaussian MAC. The rate of any set of layers is the average of its rates in these MACs over the band. Each of these rates has a closed form in which the timing offsets appear through a single number, the timing dispersion.
\item We show that for fixed spectra and offsets, the layer-rate region is a polymatroid~\cite{edmonds1970submodular,tsehanly1998polymatroid} whose vertices are attained by vector SIC over the outputs of all matched filters. Its projection onto the user rates is a pentagon. Every weighted sum rate is maximized by first decoding all layers of one user. This removes the need for an exhaustive search to check all possible permutations.
\item We prove that two layers per user, separated by half a symbol, are optimal for any delay between the users. Therefore, no coordination between the transmitters is needed. The optimal spectra follow a nested water-filling. We prove that at high SNR, the optimal density in the excess band is twice that outside it.
\item We quantify the gains numerically and implement A-RS with DVB-S2 low-density parity-check (LDPC) codes~\cite{morello2006dvbs2}. The coded link gives the weaker user a rate that no single-stream scheme can provide at the same SNR.
\end{itemize}
Section~\ref{sec:system_model} describes the system model. Section~\ref{sec:summary} summarizes the results in Fig.~\ref{fig:flow}, which lays out the structure of Sections~\ref{sec:folded_spectrum}-\ref{sec:numerical}. Section~\ref{sec:conclusion} concludes the paper.

\newcommand{\lnk}[2]{\hyperref[#2]{#1~\ref*{#2}}}
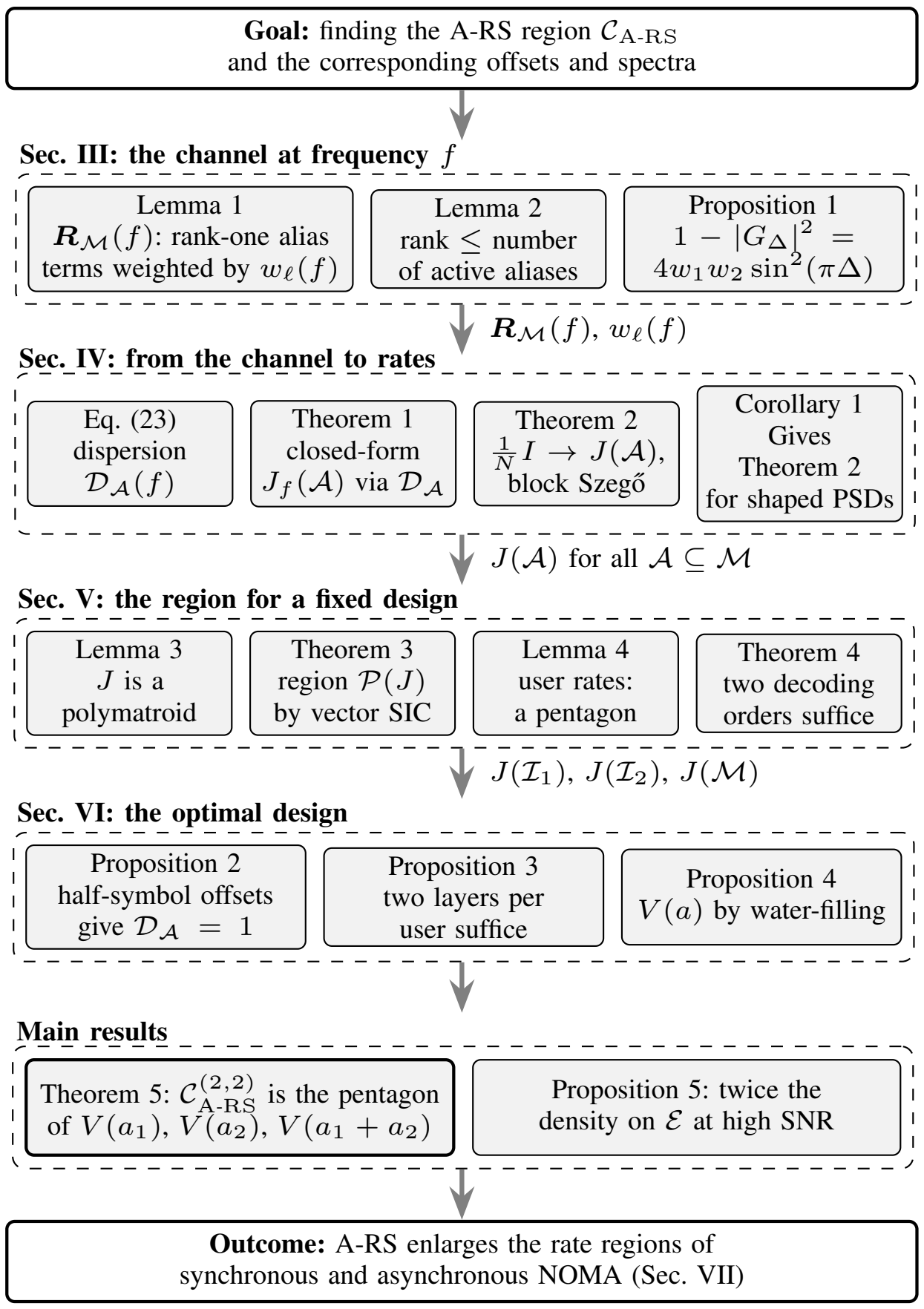
\begin{figure}[t]
\centering
\begin{tikzpicture}[font=\scriptsize, >=Stealth, node distance=0.15cm,
res/.style={draw, rounded corners=2pt, fill=gray!10, align=center, text width=#1, inner sep=2pt, minimum height=0.85cm},
res/.default=2.35cm,
stage/.style={draw, dashed, rounded corners=3pt, inner sep=3pt},
lab/.style={anchor=south west, inner sep=1pt, font=\scriptsize\bfseries},
flow/.style={->, very thick, gray}]
\node[draw, thick, rounded corners=2pt, align=center, text width=7.9cm, inner sep=3pt] (goal) {\textbf{Goal:} finding the A-RS region $\C_{\ARS}$\\ and the corresponding offsets and spectra};
\node[res=2.75cm, below=0.855cm of goal.south west, anchor=north west, xshift=0.2cm] (a1) {\lnk{Lemma}{lem:folded_spectrum}\\ $\boldsymbol R_{\mathcal M}(f)$: rank-one alias terms weighted by $w_\ell(f)$};
\node[res=1.95cm, right=of a1] (a2) {\lnk{Lemma}{lem:rank}\\ rank $\le$ number of active aliases};
\node[res, right=of a2] (a3) {\lnk{Proposition}{prop:two_alias}\\ $1-|G_\Delta|^2=4w_1w_2\sin^2(\pi\Delta)$};
\node[stage, fit=(a1)(a2)(a3)] (s1) {};
\node[lab] at (s1.north west) {\hyperref[sec:folded_spectrum]{Sec.~\ref*{sec:folded_spectrum}: the channel at frequency $f$}};
\node[res=1.685cm, below=0.855cm of s1.south west, anchor=north west, xshift=3pt] (b0) {\hyperref[eq:dispersion_def]{Eq.~(\ref*{eq:dispersion_def})}\\ dispersion $\mathcal D_{\mathcal A}(f)$};
\node[res=1.685cm, right=of b0] (b1) {\lnk{Theorem}{thm:closed_form}\\ closed-form $J_f(\mathcal A)$ via $\mathcal D_{\mathcal A}$};
\node[res=1.685cm, right=of b1] (b2) {\lnk{Theorem}{thm:decomposition}\\ $\frac1NI\to J(\mathcal A)$, block Szeg\H{o}};
\node[res=1.685cm, right=of b2] (b3) {\lnk{Corollary}{cor:shaped_decomposition}\\ Gives Theorem~\ref{thm:decomposition} for shaped PSDs};
\node[stage, fit=(b0)(b1)(b2)(b3)] (s2) {};
\node[lab] at (s2.north west) {\hyperref[sec:decomposition]{Sec.~\ref*{sec:decomposition}: from the channel to rates}};
\node[res=1.685cm, below=0.855cm of s2.south west, anchor=north west, xshift=3pt] (c1) {\lnk{Lemma}{lem:polymatroid}\\ $J$ is a polymatroid};
\node[res=1.685cm, right=of c1] (c2) {\lnk{Theorem}{thm:layer_polymatroid}\\ region $\mathcal P(J)$ by vector SIC};
\node[res=1.685cm, right=of c2] (c3) {\lnk{Lemma}{lem:aggregation}\\ user rates: a pentagon};
\node[res=1.685cm, right=of c3] (c4) {\lnk{Theorem}{thm:greedy}\\ two decoding orders suffice};
\node[stage, fit=(c1)(c2)(c3)(c4)] (s3) {};
\node[lab] at (s3.north west) {\hyperref[sec:region]{Sec.~\ref*{sec:region}: the region for a fixed design}};
\node[res, below=0.855cm of s3.south west, anchor=north west, xshift=3pt] (d1) {\lnk{Proposition}{prop:offset_optimal}\\ half-symbol offsets give $\mathcal D_{\mathcal A}=1$};
\node[res, right=of d1] (d2) {\lnk{Proposition}{prop:two_layers_enough}\\ two layers per user suffice};
\node[res, right=of d2] (d3) {\lnk{Proposition}{prop:diagonal_solution}\\ $V(a)$ by water-filling};
\node[stage, fit=(d1)(d2)(d3)] (s4) {};
\node[lab] at (s4.north west) {\hyperref[sec:design]{Sec.~\ref*{sec:design}: the optimal design}};
\node[res=3.67cm, thick, below=0.855cm of s4.south west, anchor=north west, xshift=3pt] (e1) {\lnk{Theorem}{thm:pentagon}: $\C_{\ARS}^{(2,2)}$ is the pentagon of $V(a_1)$, $V(a_2)$, $V(a_1+a_2)$};
\node[res=3.67cm, right=of e1] (e2) {\lnk{Proposition}{prop:excess_band_doubling}: twice the density on $\mathcal E$ at high SNR};
\node[stage, fit=(e1)(e2)] (s5) {};
\node[lab] at (s5.north west) {Main results};
\coordinate (ax) at (goal.south);
\draw[flow] (goal.south) -- ++(0,-0.45);
\draw[flow] (ax |- s1.south) -- node[right=3pt, black, font=\scriptsize] {$\boldsymbol R_{\mathcal M}(f)$, $w_\ell(f)$} ++(0,-0.45);
\draw[flow] (ax |- s2.south) -- node[right=3pt, black, font=\scriptsize] {$J(\mathcal A)$ for all $\mathcal A\subseteq\mathcal M$} ++(0,-0.45);
\draw[flow] (ax |- s3.south) -- node[right=3pt, black, font=\scriptsize] {$J(\mathcal I_1)$, $J(\mathcal I_2)$, $J(\mathcal M)$} ++(0,-0.45);
\draw[flow] (ax |- s4.south) -- ++(0,-0.45);
\coordinate (s5b) at (ax |- s5.south);
\node[draw, thick, rounded corners=2pt, align=center, text width=7.9cm, inner sep=3pt, below=0.45cm of s5b] (out) {\textbf{Outcome:} A-RS enlarges the rate regions of\\ synchronous and asynchronous NOMA (\hyperref[sec:numerical]{Sec.~\ref*{sec:numerical}})};
\draw[flow] (s5b) -- (out.north);
\end{tikzpicture}
\caption{Organization and connection of different entities in the paper. The first four dashed boxes are sections, and the last one collects the main results.}
\label{fig:flow}
\end{figure}

\section{System Model and Summary of Results}
\label{sec:system_model}

\subsection{Signal model}
\label{subsec:signal_model}

We consider an uplink with $K=2$ single-antenna users and one receiver over a flat-fading channel. Full channel state information (CSI) is available at both the transmitter and the receiver. Under rate splitting ~\cite{rimoldi1996,mao2022}, instead of a single codeword, User $k$ splits its message into $L_k$ independently encoded layers, Layer $\ell$ being a Gaussian codeword $\{s_{k,\ell}[n]\}$ with power $P_{k,\ell}$, where $\sum_{\ell}P_{k,\ell}=P_k$. Every layer is pulse-shaped by a common root-raised-cosine (RRC) pulse $p(t)$ with roll-off $\beta$, symbol period $T$, and occupied bandwidth $B=(1+\beta)/T$. Each transmission frame contains $N$ symbols. We normalize $\int_{\mathbb R}|p(t)|^2\mathrm{d}t=1$. Thus, $P_{k,\ell}$ and $P_k$ are energies per nominal symbol interval, called normalized powers. Each layer is transmitted with its own designed timing offset $\delta_{k,\ell}\in[0,1)$, normalized by $T$.

We collect all $M=L_1+L_2$ layers under a single index $i$, with channel $g_i=h_{k(i)}$, power $q_i=P_{k(i),\ell(i)}$, and normalized offset $\tau_i=\delta_{k(i),\ell(i)}$. Here, $k(i)$ and $\ell(i)$ denote the user and layer indices associated with $i$, respectively, and are defined as
\begin{equation}
    k(i) =
    \begin{cases}
        1, & 1\le i\le L_1\\
        2, & L_1<i\le M,
    \end{cases}
    \quad
     \ell(i) =
    \begin{cases}
        i, & 1\le i\le L_1\\
        i-L_1, & L_1<i\le M.
    \end{cases}
\end{equation}
The received waveform is
\begin{equation}\label{eq:rx_signal}
y(t)=\sum_{i=1}^{M} g_i\sqrt{q_i}\sum_{n=0}^{N-1}s_i[n]p\big(t-nT-\tau_i T\big)+\nu(t),
\end{equation}
where $\nu(t)$ is a circularly symmetric complex white Gaussian noise with power spectral density $\sigma^2$. We write $\sigma_k^2\triangleq\sigma^2/|h_k|^2$ and define the per-user signal-to-noise ratio (SNR) $a_k\triangleq P_k/\sigma_k^2$, with $a_1\ge a_2$ without loss of generality.

\begin{figure}[t]
\centering
\begin{tikzpicture}[font=\scriptsize, >=Stealth,
blk/.style={draw, rounded corners=1pt, align=center, inner sep=2pt, minimum height=0.6cm},
sum/.style={draw, circle, inner sep=0pt, minimum size=0.3cm}]
\node[blk] (m1) at (0,2.9) {$W_1$};
\node[blk] (l11) at (1.3,3.3) {$s_{1,1}[n]$, $P_{1,1}$};
\node[blk] (l12) at (1.3,2.5) {$s_{1,2}[n]$, $P_{1,2}$};
\node[blk] (p11) at (3.1,3.3) {RRC, $\delta_{1,1}T$};
\node[blk] (p12) at (3.1,2.5) {RRC, $\delta_{1,2}T$};
\node[sum] (a1) at (4.4,2.9) {$+$};
\node[blk] (m2) at (0,1.1) {$W_2$};
\node[blk] (l21) at (1.3,1.5) {$s_{2,1}[n]$, $P_{2,1}$};
\node[blk] (l22) at (1.3,0.7) {$s_{2,2}[n]$, $P_{2,2}$};
\node[blk] (p21) at (3.1,1.5) {RRC, $\delta_{2,1}T$};
\node[blk] (p22) at (3.1,0.7) {RRC, $\delta_{2,2}T$};
\node[sum] (a2) at (4.4,1.1) {$+$};
\node[sum] (c) at (6.2,2.0) {$+$};
\node (nu) at (7.3,2.0) {$\nu(t)$};
\node[blk] (mf) at (6.2,-0.5) {Matched filters\\ $p^*(-t)$ at $nT+\delta_{k,\ell}T$};
\node[blk] (sic) at (3.1,-0.5) {Vector SIC\\ (Algorithm~\ref{alg:vector_sic})};
\node (out) at (0.6,-0.5) {$\widehat W_1,\widehat W_2$};
\foreach \u in {1,2}{
\draw[->] (m\u.east) -- (l\u1.west);
\draw[->] (m\u.east) -- (l\u2.west);
\draw[->] (l\u1) -- (p\u1);
\draw[->] (l\u2) -- (p\u2);
\draw[->] (p\u1.east) -- (a\u);
\draw[->] (p\u2.east) -- (a\u);}
\draw[->] (a1) -- node[above, sloped] {$h_1$} (c);
\draw[->] (a2) -- node[above, sloped] {$h_2$} (c);
\draw[->] (nu) -- (c);
\draw[->] (c) -- node[right] {$y(t)$} (mf);
\draw[->] (mf) -- node[above] {$\boldsymbol y$} (sic);
\draw[->] (sic) -- (out);
\node[draw, dashed, rounded corners=2pt, fit=(m1)(l11)(l12)(p11)(p12)(a1), inner sep=2pt, label={[inner sep=1pt]above:User 1, $P_{1,1}+P_{1,2}=P_1$}] {};
\node[draw, dashed, rounded corners=2pt, fit=(m2)(l21)(l22)(p21)(p22)(a2), inner sep=2pt, label={[inner sep=1pt, label distance=1pt]below:User 2, $P_{2,1}+P_{2,2}=P_2$}] {};
\end{tikzpicture}
\caption{System model for $L_1=L_2=2$. User $k$ splits its message $W_k$ into Gaussian codewords $s_{k,\ell}[n]$ with powers $P_{k,\ell}$ and sends each with the RRC pulse $p(t)$ delayed by its offset $\delta_{k,\ell}T$. The receiver samples one matched filter per layer and decodes the layers by vector SIC.}
\label{fig:system}
\end{figure}
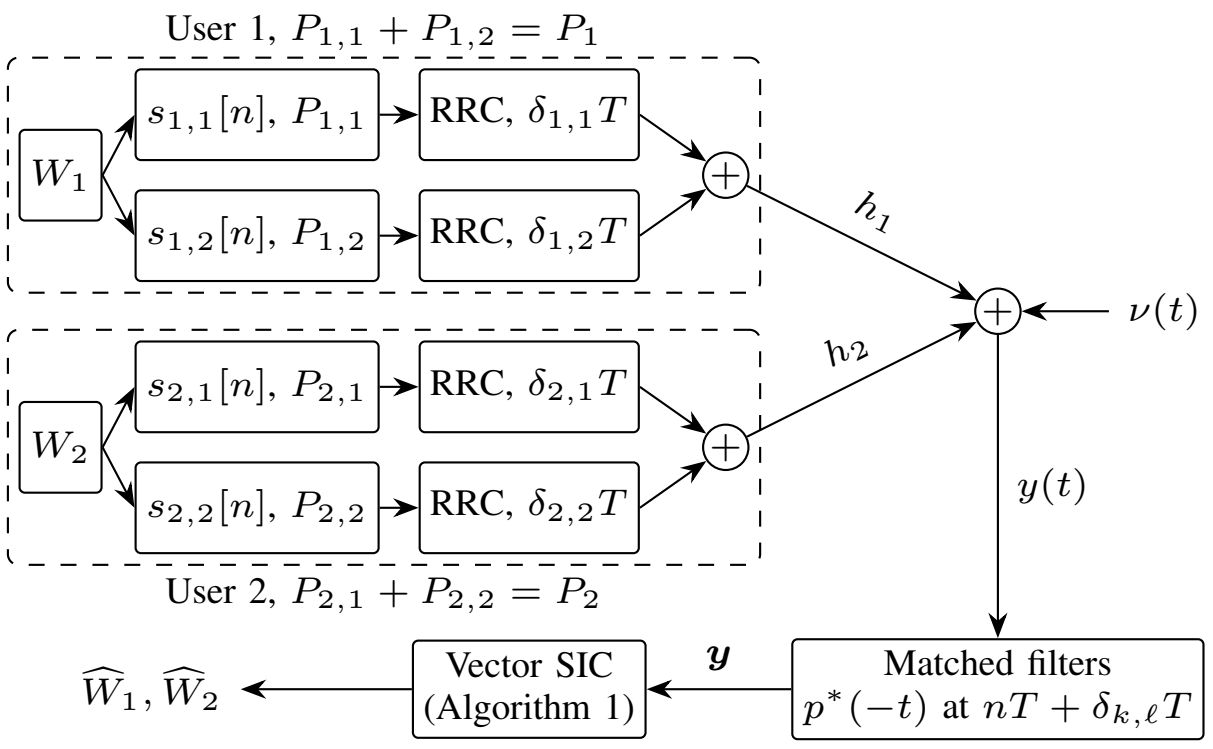

Conventional NOMA corresponds to the unsplit case $L_1=L_2=1$, in which each user transmits a single stream over the two-user asynchronous MAC. In contrast, the A-RS scheme considered here divides a user into multiple layers with distinct timing offsets. At the receiver, all $M$ layers are decoded through a single SIC chain: a layer is decoded while the not-yet-canceled layers are treated as noise, it is then subtracted, and the chain proceeds to the next layer. Fig.~\ref{fig:system} shows the transmitters and the receiver, and Fig.~\ref{fig:pulses} shows the pulse trains of the layers and the epochs at which the receiver samples them.

\newcommand{\rrcpulse}[4]{\draw[#3, shift={(#1,#2)}, xscale=1.4, yscale=0.36*#4] plot[smooth] coordinates {(-1.6,-0.030) (-1.55,-0.048) (-1.5,-0.066) (-1.45,-0.084) (-1.4,-0.102) (-1.35,-0.117) (-1.3,-0.129) (-1.25,-0.138) (-1.2,-0.142) (-1.15,-0.140) (-1.1,-0.132) (-1.05,-0.116) (-1,-0.093) (-0.95,-0.063) (-0.9,-0.024) (-0.85,0.023) (-0.8,0.077) (-0.75,0.138) (-0.7,0.205) (-0.65,0.277) (-0.6,0.352) (-0.55,0.430) (-0.5,0.509) (-0.45,0.587) (-0.4,0.662) (-0.35,0.734) (-0.3,0.799) (-0.25,0.857) (-0.2,0.907) (-0.15,0.947) (-0.1,0.976) (-0.05,0.994) (0,1.000) (0.05,0.994) (0.1,0.976) (0.15,0.947) (0.2,0.907) (0.25,0.857) (0.3,0.799) (0.35,0.734) (0.4,0.662) (0.45,0.587) (0.5,0.509) (0.55,0.430) (0.6,0.352) (0.65,0.277) (0.7,0.205) (0.75,0.138) (0.8,0.077) (0.85,0.023) (0.9,-0.024) (0.95,-0.063) (1,-0.093) (1.05,-0.116) (1.1,-0.132) (1.15,-0.140) (1.2,-0.142) (1.25,-0.138) (1.3,-0.129) (1.35,-0.117) (1.4,-0.102) (1.45,-0.084) (1.5,-0.066) (1.55,-0.048) (1.6,-0.030)};}

\begin{figure}[t]
\centering
\begin{tikzpicture}[font=\scriptsize, >=Stealth, u1/.style={black, thick}, u2/.style={black, thick, dashed}, bg/.style={gray!45}]
\begin{scope}
\clip (-0.7,-0.45) rectangle (5.0,3.4);
\foreach \y/\d/\s/\st in {2.7/0/{1,1,-1,1}/u1, 1.85/0.5/{-1,1,1,-1}/u1, 1.0/0.3/{1,1,-1,1}/u2, 0.15/0.8/{-1,1,1,-1}/u2}{
\draw[gray!30] (-0.7,\y) -- (5.0,\y);
\draw[densely dotted] (1.4+1.4*\d,\y) -- (1.4+1.4*\d,-0.45);
\foreach \a [count=\n from 0] in \s{
\ifnum\n=1 \rrcpulse{1.4*\n+1.4*\d}{\y}{\st}{\a}\else\rrcpulse{1.4*\n+1.4*\d}{\y}{bg}{\a}\fi
\fill (1.4*\n+1.4*\d,\y) circle (1.1pt);}}
\end{scope}
\node[anchor=east] at (-0.75,2.7) {$(1,1)$, $\delta_{1,1}=0$};
\node[anchor=east] at (-0.75,1.85) {$(1,2)$, $\delta_{1,2}=\frac12$};
\node[anchor=east] at (-0.75,1.0) {$(2,1)$, $\delta_{2,1}=0.3$};
\node[anchor=east] at (-0.75,0.15) {$(2,2)$, $\delta_{2,2}=0.8$};
\draw[->] (-0.7,-0.55) -- (5.05,-0.55) node[right] {$t$};
\foreach \n/\lab in {0/0, 1/T, 2/2T, 3/3T}{
\draw (1.4*\n,-0.5) -- (1.4*\n,-0.6) node[below] {$\lab$};}
\draw[<->] (1.4,2.3) -- node[above] {$T/2$} (2.1,2.3);
\end{tikzpicture}
\caption{Layer pulse trains for $L_1=L_2=2$. Each layer sends $\sum_ns_{k,\ell}[n]p(t-nT-\delta_{k,\ell}T)$, shown here with symbols $\pm1$. The pulse of $n=1$ is drawn in black, and the dots mark the sampling epochs $nT+\delta_{k,\ell}T$.}
\label{fig:pulses}
\end{figure}
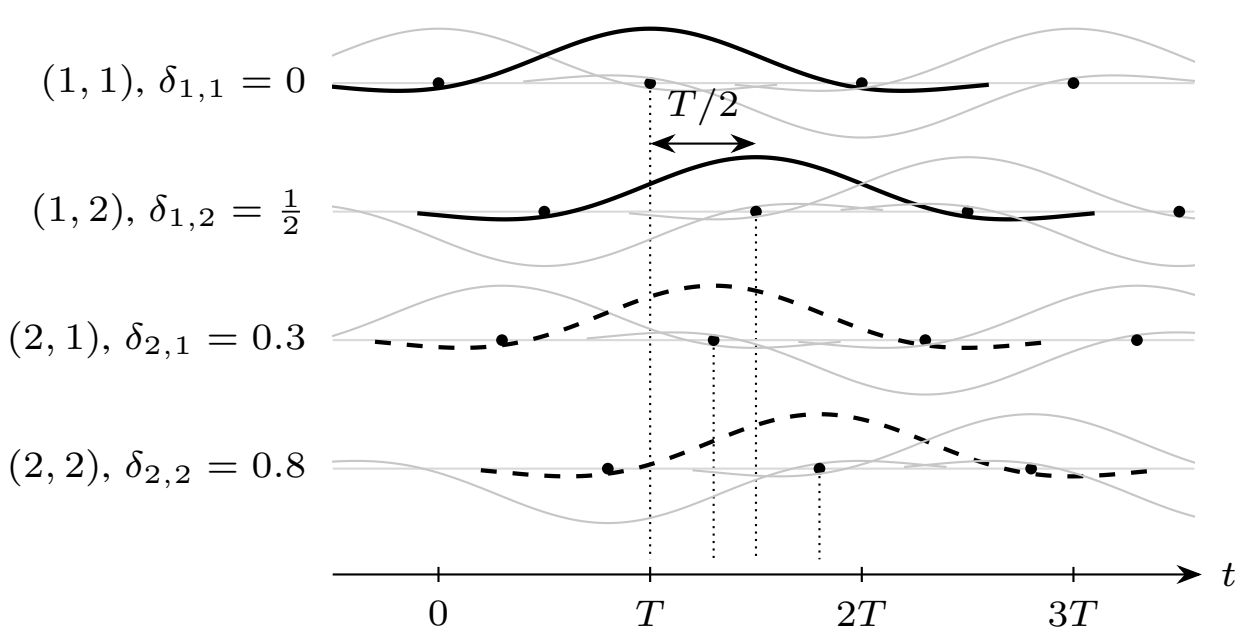

\subsection{Sufficient statistic and channel memory}
\label{subsec:sufficient_statistic}

A bank of matched filters $p^*(-t)$, one per layer, sampled at the layer epochs $\{nT+\tau_iT\}_{i=1}^{M}$, yields the sufficient statistic \cite{verdu1989asynchronous} $\boldsymbol y=[\boldsymbol y_1^{\mathsf T},\dots,\boldsymbol y_M^{\mathsf T}]^{\mathsf T}$, with
\begin{equation}\label{eq:ylayer}
\boldsymbol y_i=\sum_{j=1}^{M} g_j\sqrt{q_j}\boldsymbol \Psi_{ij}\boldsymbol s_j+\boldsymbol\nu_i,
\end{equation}
where $\boldsymbol \Psi_{ij}$ is the $N\times N$ Toeplitz matrix whose entries are samples of the pulse autocorrelation $r_p(t)=p(t)*p^*(-t)$, namely
\begin{equation}\label{eq:Rij_entries}
[\boldsymbol \Psi_{ij}]_{n,m}=r_p\big((m-n+\tau_j-\tau_i)T\big).
\end{equation}
For a square-root-Nyquist~\cite{proakis2008digital} pulse $\boldsymbol \Psi_{ii}=\boldsymbol I_N$, i.e., each layer is orthogonal to its own time shifts, whereas $\boldsymbol \Psi_{ij}=\boldsymbol \Psi_{ji}^{H}\ne\boldsymbol I_N$ whenever $\tau_i\ne\tau_j$. Distinct offsets result in a channel with memory, because each symbol of Layer $j$ overlaps two or more consecutive symbols of Layer $i$. Stacking all $M$ layers gives the equivalent discrete-time model
\begin{equation}\label{eq:stack}
\boldsymbol y=\boldsymbol \Psi\boldsymbol H\boldsymbol s+\boldsymbol\nu,
\end{equation}
where $\boldsymbol s=[\boldsymbol s_1^{\mathsf T},\dots,\boldsymbol s_M^{\mathsf T}]^{\mathsf T}\in\mathbb C^{MN}$ stacks the $M$ layer codewords, $\boldsymbol s_i=[s_i[0],\dots,s_i[N-1]]^{\mathsf T}$ collects the i.i.d.\ zero-mean unit-power Gaussian symbols of Layer $i$ so that $\E[\boldsymbol s_i\boldsymbol s_i^{H}]=\boldsymbol I_N$,
\begin{equation}\label{eq:H_layer_major}
\boldsymbol H=\diag\big(g_1\sqrt{q_1}\boldsymbol I_N,\dots,g_M\sqrt{q_M}\boldsymbol I_N\big)
\end{equation}
is the block-diagonal channel and power matrix, $\boldsymbol \Psi=[\boldsymbol \Psi_{ij}]_{i,j=1}^{M}$ is the $MN\times MN$ Gram matrix with Toeplitz blocks, and $\boldsymbol\nu=[\boldsymbol\nu_1^{\mathsf T},\dots,\boldsymbol\nu_M^{\mathsf T}]^{\mathsf T}$ is the matched-filtered noise, which is Gaussian and colored with covariance $\E[\boldsymbol\nu\boldsymbol\nu^{H}]=\sigma^2\boldsymbol \Psi$. 

\subsection{Notation}
\label{subsec:notation}

The following notation is used throughout the paper. The layer index set is
\begin{equation}\label{eq:layer_index_set}
\mathcal M\triangleq\{1,\ldots,M\},
\qquad
\mathcal I_k\triangleq\{i\in\mathcal M:k(i)=k\},
\end{equation}
so that $|\mathcal I_k|=L_k$, $\mathcal I_1\cap\mathcal I_2=\varnothing$, and $\mathcal I_1\cup\mathcal I_2=\mathcal M$. Timing offsets are always denoted by $\tau_i$ and pairwise timing differences by $\Delta_{ij}\triangleq\tau_j-\tau_i$. Every power quantity below is normalized, i.e., the value is divided by $\sigma^2$, the per-sample variance of the matched-filtered noise. Under temporally white signaling, the normalized received SNR of Layer $i$ is
\begin{equation}\label{eq:layer_snr}
\gamma_i\triangleq\frac{|g_i|^2q_i}{\sigma^2}=\frac{q_i}{\sigma_{k(i)}^2},
\qquad
\sum_{i\in\mathcal I_k}\gamma_i=\frac{P_k}{\sigma_k^2}=a_k,
\end{equation}
and, when Layer $i$ is allowed a nonflat power spectral density (PSD) $Q_i(f)\ge0$ with $\int_{\mathcal F}Q_i(f)\mathrm{d}f=q_i$, its normalized received PSD is
\begin{equation}\label{eq:rho_def}
\rho_i(f)\triangleq\frac{|g_i|^2Q_i(f)}{\sigma^2},
\qquad
\int_{\mathcal F}\rho_i(f)\mathrm{d}f=\gamma_i .
\end{equation}
The lowercase $r_i$ denotes a rate assigned to the virtual layer $i$, whereas the uppercase $R_k$ denotes the rate of the physical user $k$. The model is a complex baseband throughout the paper. 
All rates are measured accordingly in bits per complex channel use, i.e., per symbol interval of duration $T$. The normalized frequency is restricted to the fundamental interval
\begin{equation}\label{eq:fundamental_interval}
\mathcal F\triangleq\left[-\tfrac12,\tfrac12\right),
\end{equation}
and every frequency-domain quantity in this paper is periodic in $f$ with period one, so that integration over $\mathcal F$ may be replaced by integration over any interval of unit length. The continuous-time frequency, in Hz, is denoted by $F\in\R$. Finally, $\boldsymbol A\succeq\boldsymbol 0$ denotes positive semidefiniteness, $|\mathcal S|$ denotes the cardinality of a finite set and the Lebesgue measure of a measurable subset of $\R$, and $[x]^+\triangleq\max\{x,0\}$.
\subsection{Summary of Theoretical Results}
\label{sec:summary}

Our goal is finding the A-RS rate region, i.e., the set of user-rate pairs that the layered transmitter of Fig.~\ref{fig:system} can deliver over all offsets, spectra, and decoding orders, together with the design that attains its boundary. Fig.~\ref{fig:flow} shows how the results of the paper lead to this goal. Section~\ref{sec:folded_spectrum} describes the channel at a single frequency, Section~\ref{sec:decomposition} turns it into rates, Section~\ref{sec:region} turns the rates into a rate region for a fixed design, and Section~\ref{sec:design} optimizes the design.

\section{Folded-Spectrum Representation of the Layered Channel}
\label{sec:folded_spectrum}


\subsection{The correlation spectral matrix}
\label{subsec:toeplitz_symbol}

For a fixed normalized timing difference $\Delta_{ij}$, the scalar sequence generating the cross-correlation block $\boldsymbol \Psi_{ij}$ in \eqref{eq:Rij_entries} is $r_{ij}[m]=r_p((m+\Delta_{ij})T)$, $m\in\mathbb Z$. Its discrete-time Fourier transform (DTFT) is given by
\begin{equation}\label{eq:G_delta_DTFT}
G_{\Delta_{ij}}(f)
\triangleq
\sum_{m\in\mathbb Z}
r_p\big((m+\Delta_{ij})T\big)e^{-j2\pi fm},
\qquad f\in\mathcal F.
\end{equation}
The corresponding $M\times M$ correlation spectral matrix is
\begin{equation}\label{eq:R_frequency_symbol}
\boldsymbol R_{\mathcal{M}}(f)
\triangleq
\big[G_{\tau_j-\tau_i}(f)\big]_{i,j=1}^{M}.
\end{equation}
Thus, $\boldsymbol R_{\mathcal{M}}(f)$ is the DTFT of the sequence of $M\times M$ lag blocks $\boldsymbol R[m]$ with entries $[\boldsymbol R[m]]_{ij}=r_p((m+\tau_j-\tau_i)T)$.
Let $P(F)$ denote the continuous-time Fourier transform of $p(t)$, with the autocorrelation spectrum $R_p(F)=|P(F)|^2\ge0$, and define the alias weights and alias steering vectors, respectively, as
\begin{equation}\label{eq:alias_weight_def}
w_{\ell}(f)\triangleq\frac1T R_p\left(\frac{f+\ell}{T}\right)\ge0,
\qquad
\big[\boldsymbol v_{\ell}(f)\big]_i\triangleq e^{-j2\pi(f+\ell)\tau_i}.
\end{equation}

The following lemma rewrites $\boldsymbol R_{\mathcal{M}}(f)$ in terms of the pulse spectrum. In this form, $\boldsymbol R_{\mathcal{M}}(f)$ is a nonnegative combination of rank-one matrices, one for each alias frequency $F=(f+\ell)/T$ that folds onto $f$.

\begin{lemma}
\label{lem:folded_spectrum}
Let $p(t)$ be square-root Nyquist with respect to the symbol period $T$, i.e., $r_p(mT)=\delta[m]$, and let $R_p\in L^1(\R)$ (the space of absolutely integrable functions on the real line). Then, for almost every $f\in\mathcal F$,
\begin{align}
&G_{\Delta}(f)=\sum_{\ell\in\Z}w_{\ell}(f)e^{j2\pi(f+\ell)\Delta},
\label{eq:G_delta_alias_expansion}\\
&\sum_{\ell\in\Z}w_{\ell}(f)=1,
\label{eq:alias_weight_partition}\\
&\boldsymbol R_{\mathcal{M}}(f)=\sum_{\ell\in\Z}w_{\ell}(f)\boldsymbol v_{\ell}(f)\boldsymbol v_{\ell}^{H}(f).
\label{eq:R_frequency_gram_sum}
\end{align}
Consequently, $\boldsymbol R_{\mathcal{M}}(f)\succeq\boldsymbol0$, $\boldsymbol R_{\mathcal{M}}(f)$ is Hermitian with unit diagonal, $G_{-\Delta}(f)=G_{\Delta}^*(f)$, $G_0(f)=1$, and $|G_{\Delta}(f)|\le1$.
\end{lemma}

\begin{proof}
See Appendix~\ref{proof:folded_spectrum}.
\end{proof}

 By \eqref{eq:R_frequency_gram_sum}, the timing offsets affect the channel only through the phases of the alias steering vectors, and the pulse enters only through the nonnegative weights $w_{\ell}(f)$.

\subsection{Rank and the role of excess bandwidth}
\label{subsec:rank}

The rank of $\boldsymbol R_{\mathcal{M}}(f)$ is the number of signal dimensions that the $M$ layers occupy at frequency $f$. If it is one, the layers cannot be separated at $f$ by their timing offsets, as in the synchronous case. The following lemma uses \eqref{eq:R_frequency_gram_sum} to express this rank in terms of the aliases that are active at $f$.

\begin{lemma}
\label{lem:rank}
Let $\mathcal L(f)\triangleq\{\ell\in\Z:w_{\ell}(f)>0\}$ be the set of active aliases at frequency $f$. Then,
\begin{align}\label{eq:R_frequency_rank_bound}
\rank\boldsymbol R_{\mathcal{M}}(f)&=\dim\spn\{\boldsymbol v_{\ell}(f):\ell\in\mathcal L(f)\}
\\\nonumber&\le\min\big\{M,|\mathcal L(f)|\big\}.
\end{align}
In particular, if $|\mathcal L(f)|=1$, then $\boldsymbol R_{\mathcal{M}}(f)$ has rank one and $|G_{\Delta}(f)|=1$ for every $\Delta$. If all offsets coincide, i.e., $\tau_1=\cdots=\tau_M$, then $\boldsymbol R_{\mathcal{M}}(f)=\boldsymbol1_M\boldsymbol1_M^{H}$ has rank one at every frequency $f$ irrespective of the pulse.
\end{lemma}

\begin{proof}
See Appendix~\ref{proof:rank}.
\end{proof}

Lemma~\ref{lem:rank} isolates the two ingredients required by a layered asynchronous design. A single active alias, which is what a sinc pulse produces at every frequency, provides one signaling dimension no matter how many layers are transmitted. Then, changing a timing offset produces only a unit-modulus phase rotation. 
Only the combination of overlapping aliases and distinct offsets produces $\rank\boldsymbol R_{\mathcal{M}}(f)>1$, and this is the frequency-domain origin of the reported gains.

\subsection{Coupling of two overlapping aliases}
\label{subsec:two_alias}

For an RRC pulse with $0\le\beta\le1$, the support of $R_p$ is $|F|\le(1+\beta)/(2T)$. Therefore, after spectrum folding, at most two aliases are active at any frequency, and they differ by one alias index. We write
\begin{equation}\label{eq:excess_band_def}
\mathcal E\triangleq\{f\in\mathcal F:|\mathcal L(f)|=2\},
\end{equation}
for the folded excess band and note that $|\mathcal E|=\beta$. We denote the two active weights on $\mathcal E$ by $w_1(f)$ and $w_2(f)$. Eq. \eqref{eq:alias_weight_partition} is equivalent to $w_1(f)+w_2(f)=1$.

The next proposition quantifies, through $1-|G_{\Delta}(f)|^2=\det\boldsymbol R_{\mathcal{M}}(f)$ for two layers, how far $\boldsymbol R_{\mathcal{M}}(f)$ is from rank one. 

\begin{proposition}
\label{prop:two_alias}
Let two aliases $\ell_1,\ell_2$ be active at $f$ with weights $w_1(f),w_2(f)>0$. Then,
\begin{equation}\label{eq:two_alias_dimension_gain}
1-|G_{\Delta}(f)|^2=4w_1(f)w_2(f)\sin^2\big(\pi(\ell_2-\ell_1)\Delta\big).
\end{equation}
Hence, $|G_{\Delta}(f)|<1$ if and only if both $w_1(f)w_2(f)>0$ and $(\ell_2-\ell_1)\Delta\notin\Z$. For an RRC pulse, where $|\ell_2-\ell_1|=1$, the right-hand side of \eqref{eq:two_alias_dimension_gain} is equal to $4w_1(f)w_2(f)\sin^2(\pi\Delta)$ and is maximized, simultaneously at every $f\in\mathcal E$, by the half-symbol separation $\Delta=1/2$.
\end{proposition}

\begin{proof}
See Appendix~\ref{proof:two_alias}.

\end{proof}





\section{Asymptotic Per-Frequency Decomposition}
\label{sec:decomposition}

This section converts the finite $MN$-dimensional model \eqref{eq:stack} into a continuum of $M$-input memoryless Gaussian MACs indexed by frequency. The statement is asymptotic in the frame length $N$ and is the layered generalization of the symbol-asynchronous MAC decomposition~\cite{verdu1989asynchronous}. Mutual information does not change if the samples of \eqref{eq:stack} are reordered by symbol time into the slices $\widetilde{\boldsymbol y}[n]\triangleq[y_1[n],\ldots,y_M[n]]^{\mathsf T}$. Then, the offsets enter only within the $M\times M$ blocks $\boldsymbol R[r]$, not across blocks, so the Gram matrix becomes the $MN\times MN$ block-Toeplitz matrix $\boldsymbol T_N$ with the $(n,m)$-th block $\boldsymbol R[m-n]$ and the correlation spectral matrix $\sum_r\boldsymbol R[r]e^{-j2\pi fr}=\boldsymbol R_{\mathcal M}(f)$. We assume that the lag blocks are absolutely summable, $\sum_{r\in\Z}\|\boldsymbol R[r]\|_2<\infty$, which holds for RRC pulses with $\beta>0$, since their raised-cosine autocorrelation $r_p(t)$ decays as $O(|t|^{-3})$. Then, $\boldsymbol R_{\mathcal{M}}(f)$ is bounded and continuous.

\subsection{Per-frequency subset mutual information}
\label{subsec:subset_mi}

For $\mathcal A\subseteq\mathcal M$, let $\boldsymbol R_{\mathcal A}(f)$ be the principal submatrix of $\boldsymbol R_{\mathcal{M}}(f)$ indexed by $\mathcal A$ and let $\boldsymbol D_{\mathcal A}(f)\triangleq\diag\{\rho_i(f):i\in\mathcal A\}$, which reduces to $\boldsymbol D_{\mathcal A}=\diag\{\gamma_i:i\in\mathcal A\}$ for flat layer PSDs. Define
\begin{align}
J_f(\mathcal A)&\triangleq
\log_2\det\left(\boldsymbol I_{|\mathcal A|}
+\boldsymbol D_{\mathcal A}^{1/2}(f)\boldsymbol R_{\mathcal A}(f)
\boldsymbol D_{\mathcal A}^{1/2}(f)\right),
\label{eq:Jf_subset}\\
J(\mathcal A)&\triangleq\int_{\mathcal F}J_f(\mathcal A)\mathrm{d}f,
\label{eq:J_subset_integrated}
\end{align}
with $J_f(\varnothing)=J(\varnothing)=0$.
The determinant in \eqref{eq:Jf_subset} may equivalently be written as $\det(\boldsymbol I+\boldsymbol R_{\mathcal A}\boldsymbol D_{\mathcal A})$.

Define, for $\mathcal A\subseteq\mathcal M$, the aggregate normalized received PSD and the power-weighted mean timing phasor, respectively, as
\begin{equation}\label{eq:phasor_def}
\rho_{\mathcal A}(f)\triangleq\sum_{i\in\mathcal A}\rho_i(f),
\qquad
z_{\mathcal A}(f)\triangleq
\frac{1}{\rho_{\mathcal A}(f)}\sum_{i\in\mathcal A}\rho_i(f)e^{j2\pi\tau_i},
\end{equation}
and the associated timing dispersion
\begin{equation}\label{eq:dispersion_def}
\mathcal D_{\mathcal A}(f)\triangleq1-|z_{\mathcal A}(f)|^2\in[0,1].
\end{equation}
The phasor $z_{\mathcal A}$ is the power-weighted centroid of the layer offsets placed on the unit circle. Its modulus is equal to one, i.e., $|z_{\mathcal A}(f)|=1$, when all offsets in $\mathcal A$ coincide.

The log-determinant in \eqref{eq:Jf_subset} hides how $J_f(\mathcal A)$ depends on the offsets and PSDs. The next theorem reduces it to a closed form scalar .
\begin{theorem}
\label{thm:closed_form}
Let the pulse have at most two active aliases at every frequency, with the two active weights $w_1(f),w_2(f)$ satisfying $w_1(f)+w_2(f)=1$ and $w_1(f)w_2(f)=0$ outside the excess band $\mathcal E$. Then, for every $\mathcal A\subseteq\mathcal M$ and almost every $f$,
\begin{equation}\label{eq:Jf_closed_form}
J_f(\mathcal A)=\log_2\Big[1+\rho_{\mathcal A}(f)
+w_1(f)w_2(f)\rho_{\mathcal A}^2(f)\mathcal D_{\mathcal A}(f)\Big].
\end{equation}
In particular, $J_f$ depends on the timing offsets only through the single scalar $\mathcal D_{\mathcal A}(f)$, it is nondecreasing in $\mathcal D_{\mathcal A}(f)$, and it reduces to $\log_2(1+\rho_{\mathcal A}(f))$ outside $\mathcal E$ and whenever all offsets in $\mathcal A$ coincide.
\end{theorem}

\begin{proof}
See Appendix~\ref{proof:closed_form}.

\end{proof}

This theorem shows that the only way a layered design can influence a given subset rate, beyond the power it allocates, is by increasing the timing dispersion of that subset. 



\subsection{The block-Szeg\H{o} limit}
\label{subsec:szego}
So far, $J(\mathcal A)$ has been defined directly in the frequency domain. The next theorem shows that it is the per-symbol mutual information of the length-$N$ layered channel as $N\to\infty$.

\begin{theorem}
\label{thm:decomposition}
Let the layer inputs be independent, temporally white, and proper Gaussian sequences. Then, for every $\mathcal A\subseteq\mathcal M$,
\begin{equation}\label{eq:block_Szego_subset}
\lim_{N\to\infty}\frac1N
I\left(\{\boldsymbol s_i\}_{i\in\mathcal A};
\boldsymbol{y}
\mid\{\boldsymbol s_j\}_{j\notin\mathcal A}\right)
=J(\mathcal A).
\end{equation}
In particular, with $\mathcal A=\mathcal M$, the normalized mutual information of the length-$N$ layered channel converges to that of a continuum of parallel $M$-input Gaussian MACs indexed by $f\in\mathcal F$, whose frequency-$f$ member has Gram matrix $\boldsymbol R_{\mathcal{M}}(f)$ and normalized received powers $\{\rho_i(f)\}_{i\in\mathcal M}$.
\end{theorem}

\begin{proof}
See Appendix~\ref{proof:decomposition}.

\end{proof}

Theorem~\ref{thm:decomposition} expresses the limiting normalized log-determinant of an $MN\times MN$ matrix as an integral of
$M\times M$ log-determinants. It is a block version of Szeg\H{o}'s theorem~\cite{grenander1958toeplitz,gray2006toeplitz,gutierrez2008toeplitz} specialized to the layered channel. 

The following corollary extends Theorem~\ref{thm:decomposition} to shaped layer spectra.
\begin{corollary}
\label{cor:shaped_decomposition}
Let each layer $i\in\mathcal M$ transmit an independent proper Gaussian sequence with power spectral density $Q_i(f)\ge0$ that is bounded and Riemann integrable on $\mathcal F$, and let $\rho_i(f)$ be as in \eqref{eq:rho_def}. Then, under the summability assumption, for every $\mathcal A\subseteq\mathcal M$,
\begin{equation}
	\lim_{N\to\infty}
	\frac1N
	I\left(
	\{\boldsymbol s_i\}_{i\in\mathcal A};
    \boldsymbol{y}
	\mid
	\{\boldsymbol s_j\}_{j\notin\mathcal A}
	\right)
	=
	J(\mathcal A),
	\label{eq:shaped_szego}
\end{equation}
with $J(\mathcal A)$ given by \eqref{eq:J_subset_integrated} and $J_f(\mathcal A)$ by \eqref{eq:Jf_subset} with the frequency-dependent $\boldsymbol D_{\mathcal A}(f)=\diag\{\rho_i(f):i\in\mathcal A\}$.
\end{corollary}

\begin{proof}
See Appendix~\ref{proof:shaped_decomposition}.

\end{proof}

\section{Rate Region of the Layered Asynchronous MAC}
\label{sec:region}
Theorem~\ref{thm:decomposition} and Corollary~\ref{cor:shaped_decomposition} reduce the layered channel to a single set function. It shows that for every subset $\mathcal A\subseteq\mathcal M$, $J(\mathcal A)$ is the total rate that the layers in $\mathcal A$ can jointly support when the receiver already knows all the other layers. The $2^M-1$ $J(\mathcal A)$ numbers determine the layer-rate region, but as a list of constraints they are hard to deal with. Also, they do not reveal which decoder achieves the region. This section shows that $J$ has the structure that resolves both issues. 

\subsection{Polymatroid structure}
\label{subsec:polymatroid}


\begin{definition}
\label{def:polymatroid}
A set function $F:2^{\mathcal M}\to\R_+$ is a polymatroid rank function~\cite{edmonds1970submodular} if it is
\emph{normalized}, $F(\varnothing)=0$; \emph{nondecreasing},
$F(\mathcal A)\le F(\mathcal B)$ whenever
$\mathcal A\subseteq\mathcal B$; and \emph{submodular},
\begin{equation}
	F(\mathcal A)+F(\mathcal B)
	\ge
	F(\mathcal A\cup\mathcal B)+F(\mathcal A\cap\mathcal B)
	\quad
	\text{for all }
	\mathcal A,\mathcal B\subseteq\mathcal M.
	\label{eq:submodular_definition}
\end{equation}
\end{definition}
For a permutation $\pi$ of $\mathcal M$, the greedy vector $\boldsymbol r^{(\pi)}\in\R_+^M$ assigns the increment $F(\{\pi(1),\ldots,\pi(m)\})-F(\{\pi(1),\ldots,\pi(m-1)\})$ to Layer $\pi(m)$. If $F$ is a polymatroid rank function, by Edmonds' theorem~\cite{edmonds1970submodular}, the greedy vectors are exactly the vertices of the dominant face of the polytope $\{\boldsymbol r\in\R_+^M:\sum_{i\in\mathcal A}r_i\le F(\mathcal A),\ \forall\mathcal A\subseteq\mathcal M\}$.

The next lemma shows that $J$ has this structure, which makes the layer-rate region a polymatroid with SIC-achievable vertices.
\begin{lemma}
\label{lem:polymatroid}
The set function $J:2^{\mathcal M}\to\R_+$ of \eqref{eq:J_subset_integrated} is a polymatroid rank function in the sense of Definition~\ref{def:polymatroid}. That is, $J(\varnothing)=0$, $J(\mathcal A)\le J(\mathcal B)$ whenever $\mathcal A\subseteq\mathcal B$, and
\begin{equation}\label{eq:submodularity}
J(\mathcal A)+J(\mathcal B)\ge J(\mathcal A\cup\mathcal B)+J(\mathcal A\cap\mathcal B),
\end{equation}
for all $\mathcal A,\mathcal B\subseteq\mathcal M$. The same holds pointwise in $f$ for $J_f$.
\end{lemma}

\begin{proof}
See Appendix~\ref{proof:polymatroid}.

\end{proof}

\subsection{Achievable layer rates and exact SIC}
\label{subsec:sic}

Using Lemma~\ref{lem:polymatroid}, the next theorem characterizes the layer-rate region and shows that its vertices are achieved by SIC.

\begin{theorem}
\label{thm:layer_polymatroid}
Under the hypotheses of Theorem~\ref{thm:decomposition}  or of Corollary~\ref{cor:shaped_decomposition} for nonflat PSDs, the achievable layer-rate region for a fixed set of PSDs and offsets is the polymatroid
\begin{equation}\label{eq:layer_polymatroid}
\mathcal P(J)\triangleq
\left\{\boldsymbol r\in\R_+^{M}:
\sum_{i\in\mathcal A}r_i\le J(\mathcal A),
\ \forall\mathcal A\subseteq\mathcal M\right\}.
\end{equation}
Moreover, let $\pi$ be a permutation of $\mathcal M$ in which Layer $\pi(1)$ is decoded first, and let $\mathcal S_m\triangleq\{\pi(m),\ldots,\pi(M)\}$ with $\mathcal S_{M+1}\triangleq\varnothing$ denote the set of layers not yet decoded before Stage $m$. Then, the rate
\begin{equation}\label{eq:exact_sic_rate}
r_{\pi(m)}=J(\mathcal S_m)-J(\mathcal S_{m+1})
=\int_{\mathcal F}\big[J_f(\mathcal S_m)-J_f(\mathcal S_{m+1})\big]\mathrm{d}f
\end{equation}
is achievable by SIC at Stage $m$, the resulting rate vector is a vertex of $\mathcal P(J)$, and every vertex of the dominant face of $\mathcal P(J)$ arises from some $\pi$ in this way. Finally, for every $\pi$,
\begin{equation}\label{eq:sic_telescoping}
\sum_{m=1}^{M}r_{\pi(m)}=J(\mathcal M),
\end{equation}
so that the SIC sum rate does not depend on the decoding order.
\end{theorem}

\begin{proof}
See Appendix~\ref{proof:layer_polymatroid}.

\end{proof}

Note that \eqref{eq:exact_sic_rate} is the exact SIC rate and must not be replaced by a scalar SINR expression. The scalar approximation
\begin{equation}\label{eq:pairwise_approximation}
\widetilde r_{\pi(m)}
=\int_{\mathcal F}\log_2\left(1+
\tfrac{\rho_{\pi(m)}(f)}
{1+\sum_{j\in\mathcal S_{m+1}}\big|G_{\tau_{\pi(m)}-\tau_j}(f)\big|^2\rho_j(f)}\right)\mathrm{d}f
\end{equation}
is the rate of a receiver that decodes Layer $\pi(m)$ from the output of its own matched filter alone, treating the leakage $|G_{\tau_{\pi(m)}-\tau_j}(f)|^2\rho_j(f)$ of every undecoded layer as additional noise. It is generally strictly smaller than \eqref{eq:exact_sic_rate}, even for $M=2$, because the receiver discards the outputs of the other layers' matched filters. These outputs contain different noise-correlated observations of both the desired layer and the interference. 
We refer to a receiver restricted to \eqref{eq:pairwise_approximation} as a \emph{scalar-SIC} receiver, in contrast with the \emph{vector-SIC} receiver of Theorem~\ref{thm:layer_polymatroid}, which uses the outputs of all $M$ matched filters jointly.

The successive-cancellation receiver analyzed in~\cite{ganji2021rate} is of the scalar type: it treats the other user's signal as a noise with power $\eta_\tau P_2|h_2|^2$, where $\eta_\tau=\int_{\mathcal F}|G_\tau(f)|^2\mathrm{d}f$ is the coupling averaged over frequency. 
With this receiver, the two decoding orders yield different sum rates, both below the boundary of the asynchronous capacity region. This is why~\cite{ganji2021rate} finds that the maximum sum rate depends on the decoding order. Theorem~\ref{thm:layer_polymatroid} holds for any set of layers, including one layer per user. Hence, with the vector-SIC receiver, every decoding order attains a vertex of $\mathcal P(J)$ with the same sum rate $J(\mathcal M)$. Therefore, the difference comes from the different receiver design. Algorithm~\ref{alg:vector_sic} lists the transmitter and the vector-SIC receiver for a given decoding order.

\begin{algorithm}[t]
\small\DontPrintSemicolon
\caption{Layered transmission and vector SIC for a decoding order $\pi$}
\label{alg:vector_sic}
\KwIn{offsets $\{\tau_i\}$, channel gains $\{g_i\}$, normalized received layer PSDs $\{\rho_i(f)\}$, decoding order $\pi$, block length $N$, DFT frequencies $f_n=n/N$}
\KwOut{decoded layer messages and their rates $r_{\pi(1)},\ldots,r_{\pi(M)}$}
\textbf{Rates}\;
\For{$m=1,\ldots,M$}{
$\mathcal S_m\leftarrow\{\pi(m),\ldots,\pi(M)\}$\;
$r_{\pi(m)}\leftarrow J(\mathcal S_m)-J(\mathcal S_{m+1})$, by \eqref{eq:Jf_closed_form}\;
}
\textbf{Transmitter}\;
\For{each layer $i$}{
encode the message with an independent Gaussian codebook of rate below $r_i$, giving unit-variance codeword entries $X_i(f_0),\ldots,X_i(f_{N-1})$\;
$s_i[n]\leftarrow\frac{1}{\sqrt N}\sum_{k=0}^{N-1}\sqrt{Q_i(f_k)/q_i}X_i(f_k)e^{-j2\pi kn/N}$ for $n=0,\ldots,N-1$\;
User $k$ transmits $\sum_{i\in\mathcal I_k}\sqrt{q_i}\sum_ns_i[n]p(t-nT-\tau_iT)$, each layer delayed by its designed offset $\tau_iT$\;
}

\textbf{Receiver}\;
sample the matched filter $p^*(-t)$ at $nT+\tau_iT$ for every layer, divide by $\sigma$, and invert the transmitter transform, $\boldsymbol Y(f_n)\leftarrow\frac{1}{\sqrt N}\sum_{m=0}^{N-1}\widetilde{\boldsymbol y}[m]e^{j2\pi nm/N}\in\mathbb C^M$\;
let the columns of $\boldsymbol V(f_n)$ be the steering vectors $\boldsymbol v_\ell(f_n)$ of the active aliases, two in $\mathcal E$ and one outside, and $\boldsymbol W(f_n)=\diag\{w_\ell(f_n)\}$\;
project onto the active aliases and whiten the noise: $\boldsymbol Z(f_n)\leftarrow\boldsymbol W^{-1/2}(f_n)\big(\boldsymbol V^H(f_n)\boldsymbol V(f_n)\big)^{-1}\boldsymbol V^H(f_n)\boldsymbol Y(f_n)$\;
signature of Layer $i$ in $\boldsymbol Z(f_n)$: $\boldsymbol b_i(f_n)\leftarrow\frac{g_i}{|g_i|}\sqrt{\rho_i(f_n)}\boldsymbol W^{1/2}(f_n)\boldsymbol V^H(f_n)\boldsymbol e_i$ for every layer $i$\;
\For{$m=1,\ldots,M$}{
layer decoded at Stage $m$: $u\leftarrow\pi(m)$\;
\For{$n=0,\ldots,N-1$}{
covariance of the noise plus the layers not yet decoded: $\boldsymbol C_n\leftarrow\boldsymbol I+\sum_{j\in\mathcal S_{m+1}}\boldsymbol b_j(f_n)\boldsymbol b_j^H(f_n)$\;
SINR at the output of the MMSE filter: $\mathrm{SINR}_n\leftarrow\boldsymbol b_u^H(f_n)\boldsymbol C_n^{-1}\boldsymbol b_u(f_n)$\;
unbiased MMSE estimate of the codeword entry: $\widehat X_u(f_n)\leftarrow\boldsymbol b_u^H(f_n)\boldsymbol C_n^{-1}\boldsymbol Z(f_n)/\mathrm{SINR}_n$\;
}
decode Layer $u$ from $\{\widehat X_u(f_n)\}$, treating bin $n$ as an AWGN channel with noise variance $1/\mathrm{SINR}_n$\;
re-encode Layer $u$ and cancel it: $\boldsymbol Z(f_n)\leftarrow\boldsymbol Z(f_n)-\boldsymbol b_u(f_n)X_u(f_n)$ for every $n$\;
}
\end{algorithm}

\subsection{Projection onto physical-user rates}
\label{subsec:projection}

The physical rate of User $k$ is the sum of the rates of its layers, i.e., $R_k=\sum_{i\in\mathcal I_k}r_i$. The following lemma shows that this aggregation preserves the polymatroid structure, so that the two-user region is described by only three inequalities.

\begin{lemma}
\label{lem:aggregation}
Let $J$ be a polymatroid rank function on $\mathcal M$ and let $\{\mathcal I_1,\mathcal I_2\}$ be a partition of $\mathcal M$. Then,
\begin{align}
&\left\{(R_1,R_2)\in\R_+^2:\exists\boldsymbol r\in\mathcal P(J),
R_k=\textstyle\sum_{i\in\mathcal I_k}r_i\right\}
\nonumber\\
&\quad=
\left\{(R_1,R_2)\in\R_+^2:
\begin{array}{l}
R_1\le J(\mathcal I_1),\ R_2\le J(\mathcal I_2),\\
R_1+R_2\le J(\mathcal M)
\end{array}
\right\}.
\label{eq:aggregated_region}
\end{align}
\end{lemma}

\begin{proof}
See Appendix~\ref{proof:aggregation}.

\end{proof}

\subsection{Weighted boundary without order enumeration}
\label{subsec:greedy}

A convex rate region is completely described by its weighted sum rates. 
For a fixed design, a direct computation would compare the $M!$ SIC orders of Theorem~\ref{thm:layer_polymatroid}. The following theorem shows that this is unnecessary: the maximum has a closed form in terms of $J(\mathcal I_1)$, $J(\mathcal I_2)$, and $J(\mathcal M)$. The maximum is achieved by decoding all layers of one user before those of the other.
\begin{theorem}
\label{thm:greedy}
For a compact convex region $\mathcal C\subset\R_+^2$ and a weight vector $\boldsymbol\mu=(\mu_1,\mu_2)$ with $\mu_k\ge0$ and $\mu_1+\mu_2=1$, define
\begin{equation}\label{eq:support}
	h_{\mathcal C}(\boldsymbol\mu)\triangleq
	\max_{(R_1,R_2)\in\mathcal C}\mu_1R_1+\mu_2R_2.
\end{equation}
For a fixed set of layer PSDs and timing offsets, let $\mathcal C$ denote the aggregated two-user region \eqref{eq:aggregated_region}. Then,
\begin{equation}\label{eq:weighted_objective}
h_{\mathcal C}(\boldsymbol\mu)=
\begin{cases}
\mu_2J(\mathcal M)+(\mu_1-\mu_2)J(\mathcal I_1),&\mu_1\ge\mu_2,\\
\mu_1J(\mathcal M)+(\mu_2-\mu_1)J(\mathcal I_2),&\mu_2\ge\mu_1.
\end{cases}
\end{equation}
The maximum is achieved by any SIC order that decodes all layers of the lower-weight user before all layers of the higher-weight user. The order among the layers of the same physical user does not affect $h_{\mathcal C}(\boldsymbol\mu)$.
\end{theorem}

\begin{proof}
See Appendix~\ref{proof:greedy}.

\end{proof}


\subsection{The layered asynchronous rate-splitting region}
\label{subsec:region_def}

For fixed layer counts $(L_1,L_2)$, the layered asynchronous rate-splitting region is
\begin{align}
\C_{\ARS}^{(L_1,L_2)}\triangleq
\overline{\conv}\bigcup_{\{Q_i\},\{\tau_i\},\pi}
{}&\Bigg\{(R_1,R_2):R_k=\sum_{i\in\mathcal I_k}r_i,
\nonumber\\
&\sum_{i\in\mathcal I_k}\int_{\mathcal F}Q_i(f)\mathrm{d}f\le P_k\Bigg\},
\label{eq:ARS_region}
\end{align}
where the union is over all nonnegative layer PSDs, all offset vectors $\{\tau_i\}\in[0,1)^M$, and all SIC orders, and the closed convex hull accounts for time sharing. Here, $\operatorname{conv}(\mathcal S)$ denotes the convex hull of a set $\mathcal S$, and $\overline{\operatorname{conv}}(\mathcal S)$ denotes its closure.
For a fixed offset vector and fixed PSDs, Lemma~\ref{lem:aggregation} shows that the inner set in \eqref{eq:ARS_region} is a pentagon. The union over PSDs and offsets of these pentagons is in general not a pentagon. This is the mechanism that produces a curved dominant face~\cite{ganji2021rate, verdu1989asynchronous}, the part of the boundary on which no rate pair can be improved in one coordinate without giving up some of the other. 
For a fixed allocation, SIC is optimal by Theorem~\ref{thm:layer_polymatroid}. Therefore, the curvature is a property of the transmitter, not of the receiver. It appears when the allocation that maximizes $J(\mathcal I_k)$ differs from the one that maximizes $J(\mathcal M)$. Theorem~\ref{thm:pentagon} shows that this does not happen for A-RS.


\section{Design and Optimization}
\label{sec:design}

\subsection{The weighted design problem}
\label{subsec:design_problem}

By Theorem~\ref{thm:greedy}, the SIC order is fixed by the weights. Let $\mathcal C(\{\tau_i\},\{Q_i\})$ denote the region \eqref{eq:aggregated_region} for the offsets $\{\tau_i\}$ and PSDs $\{Q_i\}$. For each weight $\boldsymbol\mu$, tracing the boundary of $\C_{\ARS}^{(L_1,L_2)}$ requires
\begin{align}
h_{\ARS}^{(L_1,L_2)}(\boldsymbol\mu)=
&\max_{\{\tau_i\},\{Q_i(f)\ge0\}}\quad
h_{\mathcal C(\{\tau_i\},\{Q_i\})}(\boldsymbol\mu)
\label{eq:joint_problem}\\
\text{s.t.}\quad
&\sum_{i\in\mathcal I_k}\int_{\mathcal F}Q_i(f)\mathrm{d}f\le P_k,
\quad k\in\{1,2\},
\nonumber
\end{align}
where the objective is given in closed form by \eqref{eq:weighted_objective}.
Problem \eqref{eq:joint_problem} is nonconvex in the offsets. We show that, with two layers per user, the optimal offsets are known in closed form. As a result, the remaining power allocation reduces to a scalar water-filling problem.

\subsection{Optimal timing offsets and number of layers}
\label{subsec:offsets}

Since $J_f(\mathcal A)$ is nondecreasing in $\mathcal D_{\mathcal A}(f)$ (Theorem~\ref{thm:closed_form}) and $\mathcal D_{\mathcal A}(f)\leq 1$, as defined in \eqref{eq:dispersion_def}, offsets that make every dispersion equal to one are optimal. The next proposition provides such a design. In this section, $\rho_k(f)\triangleq\rho_{\mathcal I_k}(f)$ denotes the aggregate spectrum of User $k$.

\begin{proposition}
\label{prop:offset_optimal}
Let $L_1=L_2=2$ and let each user's two layers be separated by half a symbol, $\delta_{k,2}=\delta_{k,1}+1/2$, with an equal intra-user power split at every frequency. Then
\begin{equation}\label{eq:all_dispersions_one}
\mathcal D_{\mathcal I_1}(f)=\mathcal D_{\mathcal I_2}(f)=\mathcal D_{\mathcal M}(f)=1,
\end{equation}
for almost every $f$, for any inter-user delay and spectra $\rho_1(f)$ and $\rho_2(f)$. Consequently, this design maximizes $h_{\ARS}^{(2,2)}(\boldsymbol\mu)$ for every $\boldsymbol\mu$ over all offsets and all intra-user power splits.
\end{proposition}

\begin{proof}
With $\delta_{k,2}=\delta_{k,1}+1/2$ and equal intra-user powers,
\begin{equation}\label{eq:user_phasor_vanishes}
\rho_{\mathcal I_k}(f)z_{\mathcal I_k}(f)=\frac{\rho_k(f)}{2}e^{j2\pi\delta_{k,1}}\left(1+e^{j\pi}\right)=0 .
\end{equation}
Hence, $z_{\mathcal I_1}=z_{\mathcal I_2}=0$, and since $\rho_{\mathcal M}z_{\mathcal M}=\rho_{\mathcal I_1}z_{\mathcal I_1}+\rho_{\mathcal I_2}z_{\mathcal I_2}$, also $z_{\mathcal M}=0$. By Theorem~\ref{thm:closed_form}, each $J(\mathcal A)$ is nondecreasing in $\mathcal D_{\mathcal A}\le1$, so $J(\mathcal I_1)$, $J(\mathcal I_2)$ and $J(\mathcal M)$ are maximized simultaneously, and so is every branch of \eqref{eq:weighted_objective}.
\end{proof}

Therefore, the relative delay between users is immaterial and no timing coordination between the transmitters is needed, provided that each transmitter separates its own two layers by half a symbol.

\begin{proposition}
\label{prop:two_layers_enough}
Let the pulse be RRC with $0\le\beta\le1$. Then, $\C_{\ARS}^{(L_1,L_2)}\subseteq\C_{\ARS}^{(2,2)}$ for all $L_1,L_2\ge1$, with equality when $L_1,L_2\ge2$.
\end{proposition}

\begin{proof}
Fix any layer counts, offsets, and PSDs. By Lemma~\ref{lem:aggregation}, the achievable set is the pentagon of the triple $\big(J(\mathcal I_1),J(\mathcal I_2),J(\mathcal M)\big)$, and by Theorem~\ref{thm:closed_form} each entry depends only on the corresponding aggregate spectrum and is nondecreasing in its dispersion. The $(2,2)$ design of Proposition~\ref{prop:offset_optimal} with the same per-user aggregate spectra uses the same power and has all dispersions equal to one, so its triple, and hence its pentagon, is at least as large. The equality for $L_k\ge2$ follows by giving the extra layers zero power.
\end{proof}

\subsection{Optimal power allocation}
\label{subsec:power_allocation}

It remains to choose the aggregate spectra $\rho_1(f)$ and $\rho_2(f)$. Let $w(f)\triangleq w_1(f)w_2(f)\in[0,1/4]$ and $\rho_{\mathcal M}(f)=\rho_1(f)+\rho_2(f)$. Since all dispersions are equal to one, Theorem~\ref{thm:closed_form} gives
\begin{align}
J(\mathcal I_k)&=\int_{\mathcal F}\log_2\big(1+\rho_k(f)+w(f)\rho_k^2(f)\big)\mathrm{d}f,
\nonumber\\
J(\mathcal M)&=\int_{\mathcal F}\log_2\big(1+\rho_{\mathcal M}(f)+w(f)\rho_{\mathcal M}^2(f)\big)\mathrm{d}f
\label{eq:J_aggregate_only}
\end{align}
Therefore, all three subset rates are the same functional evaluated at different aggregate spectra. As a result, each subset rate is bounded by the solution of the scalar problem
\begin{equation}
V(a)\triangleq
\max_{\substack{\rho(f)\ge0\\ \int_{\mathcal F}\rho(f)\mathrm{d}f\le a}}
\int_{\mathcal F}\log_2\big(1+\rho(f)+w(f)\rho^2(f)\big)\mathrm{d}f .
\label{eq:V_def}
\end{equation}

\begin{proposition}
\label{prop:diagonal_solution}
For $a>0$, the maximizer of \eqref{eq:V_def} is unique up to null sets and is
\begin{equation}\label{eq:diagonal_kkt}
\rho^{\star}(f)=
\begin{cases}
\left[\tfrac{2w(f)-\nu+\sqrt{\nu^2+4w^2(f)-4\nu^2w(f)}}
{2\nu w(f)}\right]^{+},&w(f)>0,\\[2ex]
\left[\tfrac1\nu-1\right]^{+},&w(f)=0,
\end{cases}
\end{equation}
where $\nu\in(0,1)$ is the unique value for which $\int_{\mathcal F}\rho^\star(f)\mathrm{d}f=a$.
\end{proposition}

\begin{proof}
See Appendix~\ref{proof:diagonal_solution}.

\end{proof}

The spectrum is thus computed exactly as in water-filling: for a trial $\nu$, evaluate \eqref{eq:diagonal_kkt} at every frequency, compare the resulting power with $a$, and update $\nu$ by bisection.

\begin{lemma}\label{lem:monotone_spectrum}
If $0\le a\le b$, then $\rho_a^\star(f)\le\rho_b^\star(f)$ almost everywhere.
\end{lemma}
\begin{proof}
A larger budget requires a smaller $\nu$, and the root of $g_f(\rho(f))=\nu$, with $g_f$ as in Appendix~\ref{proof:diagonal_solution}, increases as $\nu$ decreases.
\end{proof}

\begin{theorem}
\label{thm:pentagon}
Let $L_1=L_2=2$. Then
\begin{equation}\label{eq:pentagon}
\C_{\ARS}^{(2,2)}=\left\{(R_1,R_2)\in\R_+^2:
\begin{array}{l}
R_k\le V(a_k),\ k\in\{1,2\},\\
R_1+R_2\le V(a_1+a_2)
\end{array}\right\},
\end{equation}
and every boundary point is attained with intra-user offsets $\{0,1/2\}$ and an equal intra-user power split. In particular, its dominant face is the straight segment $R_1+R_2=V(a_1+a_2)$.
\end{theorem}

\begin{proof}
For any offsets and PSDs, Theorem~\ref{thm:closed_form} and $\mathcal D_{\mathcal A}(f)\le1$ gives $J(\mathcal I_k)\le\int_{\mathcal F}\log_2\big(1+\rho_k(f)+w(f)\rho_k^2(f)\big)\mathrm{d}f\le V(a_k)$. In the same way, $J(\mathcal M)\le V(a_1+a_2)$, which with Lemma~\ref{lem:aggregation} proves $\subseteq$. For $\supseteq$, give User 1 the spectrum $\rho_1(f)=\rho_{a_1}^{\star}(f)$ and User 2 the spectrum $\rho_2(f)=\rho_{a_1+a_2}^{\star}(f)-\rho_{a_1}^{\star}(f)$, which is nonnegative by Lemma~\ref{lem:monotone_spectrum} and integrates to $a_2$. Then, $\rho_{\mathcal M}(f)=\rho_{a_1+a_2}^{\star}(f)$. Therefore, this allocation attains $J(\mathcal I_1)=V(a_1)$ and $J(\mathcal M)=V(a_1+a_2)$ simultaneously and provides the corner $\big(V(a_1),V(a_1+a_2)-V(a_1)\big)$. Switching users in the argument provides the other corner, and time sharing fills the segment between them.
\end{proof}

Therefore, the complete design is a nested water-filling. Users jointly occupy $\rho_{a_1+a_2}^\star$. The higher-weight user takes the inner part $\rho_{a_k}^\star$ and is decoded last. Each user's spectrum is split equally between its two half-symbol-separated layers. For a single user, the theorem gives the rate $V(a_k)$.

Theorem~\ref{thm:pentagon} contrasts with the known regions of asynchronous and frequency-selective multiple-access channels with one stream per user, whose dominant face is curved~\cite{verdu1989asynchronous,chengverdu1993waterfilling,ganji2021rate}. The curvature arises because the sum rate depends on how the two users share each frequency. For the asynchronous channel, the sharing is done through dispersion $\mathcal D_{\mathcal M}(f)$.
With two half-symbol-separated layers per user, all dispersions are equal to one for every allocation. Thus, the union of pentagons collapses to a single pentagon and SIC with time sharing achieves every boundary point.

\subsection{High-SNR behavior of the optimal spectrum}
\label{subsec:high_snr}
The next proposition shows how the optimal spectrum uses the excess band at high SNR and compares $V(a)$ with the band-limited benchmark.

\begin{proposition}
\label{prop:excess_band_doubling}
Let the pulse be RRC with $\beta\in(0,1]$. As $a\to\infty$,
\begin{equation}
\frac{\rho_a^\star(f)}{a}
\longrightarrow
\begin{cases}
\dfrac{2}{1+\beta},&f\in\mathcal E,\\[1ex]
\dfrac{1}{1+\beta},&f\notin\mathcal E,
\end{cases}
\label{eq:doubling}
\end{equation}
almost everywhere, so the optimal power density inside the folded excess band is asymptotically twice that outside it. Moreover,
\begin{equation}
V(a)=C_{\mathrm{BL}}(a)-2\beta+o(1),
\label{eq:V_asymptote}
\end{equation}
where $C_{\mathrm{BL}}(a)\triangleq(1+\beta)\log_2\big(1+a/(1+\beta)\big)$ is the band-limited benchmark.
\end{proposition}

\begin{proof}
See Appendix~\ref{proof:excess_band_doubling}.

\end{proof}

With a flat aggregate PSD $\rho(f)=a$, the same expansion gives $(1+\beta)\log_2a-4\beta+o(1)$. Its asymptotic gap to $C_{\mathrm{BL}}$ is $4\beta-(1+\beta)\log_2(1+\beta)$, and spectrum optimization gains $2\beta-(1+\beta)\log_2(1+\beta)$ bits per nominal symbol. This gain is concave in $\beta$, vanishes at $\beta=0$ and $\beta=1$, and is largest at $\beta=4/e-1\approx0.472$, where it is about $0.12$ bits.

\section{Numerical Results}
\label{sec:numerical}

The results are illustrated for the RRC pulse with $\beta=0.5$ and the integrals over $\mathcal F$ are evaluated as means over a uniform grid with $N_f=1024$ frequency points. Rates are in bits per symbol interval $T$. Using the same per-user SNRs $a_1$ and $a_2$, three systems are compared: (i) synchronous NOMA ($L_1=L_2=1$, $\tau_1=\tau_2$), whose region is the pentagon with sides $\log_2(1+a_1)$, $\log_2(1+a_2)$ and $\log_2(1+a_1+a_2)$, (ii) asynchronous NOMA (A-NOMA, $L_1=L_2=1$ with inter-user delay $1/2$), and (iii)A-RS with $L_1=L_2=2$ and the design of Proposition~\ref{prop:offset_optimal}. Each asynchronous scheme is evaluated with flat PSDs, $\rho_i(f)=\gamma_i$, and with optimized PSDs. The optimized A-RS region is the pentagon of Theorem~\ref{thm:pentagon}, with $V(\cdot)$ calculated using bisection on $\nu$ in \eqref{eq:diagonal_kkt}.

\begin{figure}[t]
\centering
\begin{subfigure}{\columnwidth}
\centering
\includegraphics[width=\columnwidth]{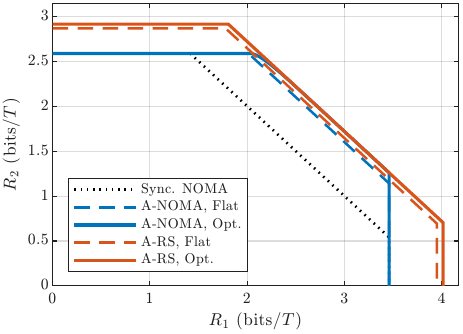}
\caption{$(a_1,a_2)=(10,5)$}
\label{fig:regions_a}
\end{subfigure}

\begin{subfigure}{\columnwidth}
\centering
\includegraphics[width=\columnwidth]{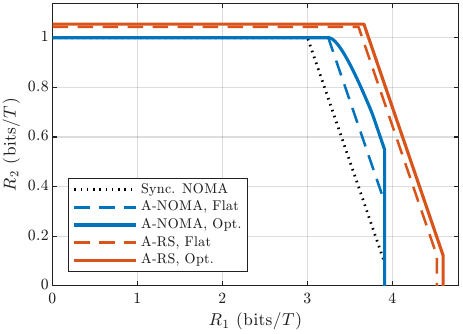}
\caption{$(a_1,a_2)=(14,1)$}
\label{fig:regions_b}
\end{subfigure}
\caption{Rate regions for $\beta=0.5$. Dashed lines shows the flat PSDs. Solid lines show optimized PSDs. The optimized A-RS region is the pentagon of Theorem~\ref{thm:pentagon}.}
\label{fig:regions}
\end{figure}

Fig.~\ref{fig:regions} shows the regions for $(a_1,a_2)=(10,5)$ and $(14,1)$, which have the same total SNR. A single layer has $\mathcal D_{\mathcal I_k}(f)=0$. Therefore, by Theorem~\ref{thm:closed_form}, A-NOMA cannot give User $k$ more than the synchronous rate of $\log_2(1+a_k)$, independent of its spectrum. A-RS increases this rate to $V(a_k)$, from $3.459$ to $4.015$ for $a_1=10$ and from $2.585$ to $2.913$ for $a_2=5$, because the two layers of a user exploit the second dimension available on $\mathcal E$.

Synchronous NOMA has the sum rate $\log_2 16=4$ in both cases. With flat PSDs, A-NOMA reaches $4.600$ for $(10,5)$ and $4.244$ for $(14,1)$, whereas A-RS reaches $4.644$ for both, since its sum rate depends only on $a_1+a_2$. With optimized PSDs, A-RS reaches $V(15)=4.718$ for both. Since $4\rho_1(f)\rho_2(f)\le\rho_{\mathcal M}^2(f)$ with equality if and only if $\rho_1(f)=\rho_2(f)$, A-NOMA achieves $V(a_1+a_2)$ if and only if each user can supply half of the power of $\rho_{a_1+a_2}^\star(f)$ on $\mathcal E$, i.e.,
\begin{equation}
\min\{a_1,a_2\}\ge\frac12\int_{\mathcal E}\rho_{a_1+a_2}^\star(f)\mathrm{d}f,
\label{eq:noma_sum_condition}
\end{equation}
where the right-hand side is equal to $4.710$ for $a_1+a_2=15$. Condition \eqref{eq:noma_sum_condition} holds for $(10,5)$, where the two optimized sum rates coincide, and fails for $(14,1)$, where A-NOMA reaches $4.456$ and A-RS gains $0.262$ bits.

The optimized A-NOMA region has a curved boundary, because the spectra that maximize $J(\mathcal I_k)$ and $J(\mathcal M)$ differ, whereas the optimized A-RS region is a pentagon. Even when the maximum sum rates coincide, the A-RS region strictly contains the A-NOMA region.

\begin{figure}[t]
\centering
\begin{subfigure}{\columnwidth}
\centering
\includegraphics[width=\columnwidth]{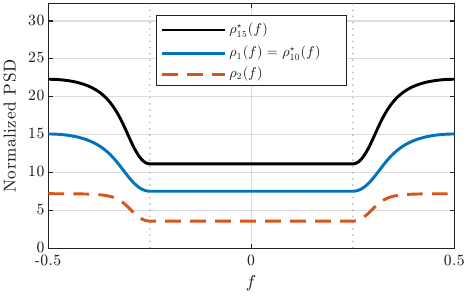}
\caption{Water-filling, $(a_1,a_2)=(10,5)$, with User 1 decoded last}
\label{fig:spectrum_a}
\end{subfigure}

\begin{subfigure}{\columnwidth}
\centering
\includegraphics[width=\columnwidth]{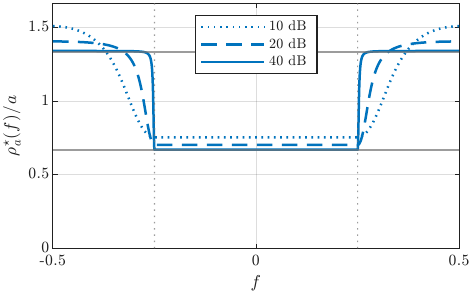}
\caption{Gray lines show the limits $2/(1+\beta)$ and $1/(1+\beta)$ of \eqref{eq:doubling}}
\label{fig:spectrum_b}
\end{subfigure}

\begin{subfigure}{\columnwidth}
\centering
\includegraphics[width=\columnwidth]{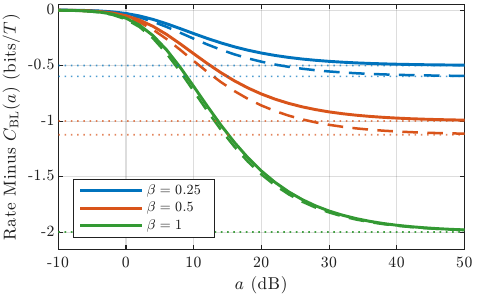}
\caption{Gap of $V(a)$ (solid) and of the flat-PSD rate (dashed) to $C_{\mathrm{BL}}(a)$. The dotted lines show the asymptotes $-2\beta$ and $-4\beta+(1+\beta)\log_2(1+\beta)$}
\label{fig:spectrum_c}
\end{subfigure}
\caption{Optimal spectra. The vertical dotted lines in (a) and (b) mark the edges of $\mathcal E$.}
\label{fig:spectrum}
\end{figure}

Fig.~\ref{fig:spectrum}(a) shows the nested water-filling of Theorem~\ref{thm:pentagon} at $(a_1,a_2)=(10,5)$ for the corner at which User 1 is decoded last. User 1 transmits $\rho_{10}^\star(f)$, User 2 transmits $\rho_{15}^\star(f)-\rho_{10}^\star(f)\ge0$, and each spectrum is split equally between the two layers of its user. For the RRC pulse, $\mathcal E=\{f\in\mathcal F:|f|\ge(1-\beta)/2\}$, which is $0.25\le|f|<0.5$ for $\beta=0.5$. All spectra have a higher density on $\mathcal E$, where two aliases are active. Fig.~\ref{fig:spectrum}(b) shows that $\rho_a^\star(f)/a$ approaches $2/(1+\beta)=1.333$ on $\mathcal E$ and $1/(1+\beta)=0.667$ outside it, as in \eqref{eq:doubling}. 

Fig.~\ref{fig:spectrum}(c) shows the gaps of $V(a)$ and of the flat-PSD rate to $C_{\mathrm{BL}}(a)$. At low SNR, both gaps vanish, since all three rates approach $a\log_2e$. At high SNR, they approach $-2\beta$ and $-4\beta+(1+\beta)\log_2(1+\beta)$, as stated by Proposition~\ref{prop:excess_band_doubling}. At $50$ dB, spectrum optimization gains $0.097$, $0.122$, and $0.001$ bits for $\beta=0.25$, $0.5$, and $1$, respectively, close to the limits $0.098$, $0.123$, and $0$ of $2\beta-(1+\beta)\log_2(1+\beta)$.

\subsection{A coded implementation}
\label{subsec:num_coded}

We finally implement A-RS with practical codes for $\beta=1$ and $a_1=2a_2$. Each user transmits two layers at offsets $\{0,1/2\}$ with equal powers, and the users are aligned on a common half-symbol grid, which does not impose any restriction on the system according Proposition~\ref{prop:offset_optimal}. Each layer is encoded by DVB-S2 LDPC codes of length $64800$~\cite{morello2006dvbs2}, whose bits are interleaved and Gray-mapped onto the constellation as in bit-interleaved coded modulation~\cite{caire1998bicm} and DFT-precoded as in Algorithm~\ref{alg:vector_sic} with a flat PSD. User 1 utilizes QPSK with rate $1/3$ in both layers, resulting in $R_1=4/3$. User 2 utilizes 16QAM with rate $4/5$ and 64QAM with rate $3/4$, resulting in $R_2=3.2+4.5=7.7$ bits. The receiver is the vector SIC of Algorithm~\ref{alg:vector_sic}, which decodes the two layers of User 1, then the 16QAM layer, and finally the 64QAM layer. The demapper computes max-log bit LLRs, treating the undecoded layers as Gaussian noise, and the LDPC decoder runs at most $200$ iterations. 
A frame fails if any decoded  bit from any layer is wrong.

\begin{figure}[t]
\centering
\begin{subfigure}{\columnwidth}
\centering
\includegraphics[width=\columnwidth]{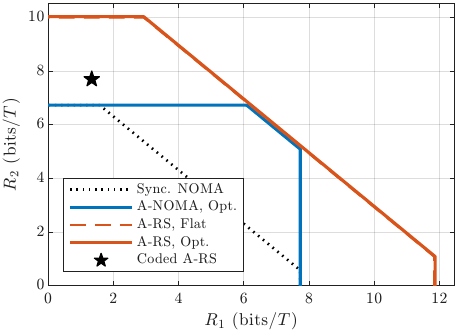}
\caption{Gaussian regions at $a_1+a_2=25$ dB and the offered rate pair $(R_1,R_2)=(4/3,7.7)$}
\label{fig:coded_a}
\end{subfigure}

\begin{subfigure}{\columnwidth}
\centering
\includegraphics[width=\columnwidth]{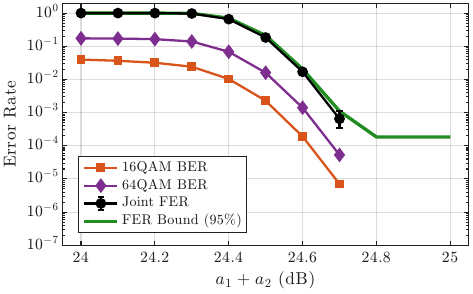}
\caption{Bit-error rate of the 16QAM and 64QAM layers and joint frame-error rate with $95\%$ confidence intervals. The green line shows the $95\%$ upper bounds after zero failures in $20000$ frames}
\label{fig:coded_b}
\end{subfigure}
\caption{LDPC-coded A-RS link, $\beta=1$, $a_1=2a_2$.}
\label{fig:coded}
\end{figure}

For each SNR, frames are sent until $100$ frames fail or $20000$ frames are sent, and the frame-error rate (FER) is the fraction of failed frames. The number of failed frames is binomial, and Fig.~\ref{fig:coded}(b) shows its exact $95\%$ confidence intervals. If no frame fails in $n$ frames, the upper end of this interval is the FER $p$ at which zero failures still have probability $(1-p)^n=0.025$, the upper $2.5\%$ tail of the $95\%$ interval, i.e., $1.8\times10^{-4}$ for $n=20000$. At $24.2$ dB, every frame fails; at $24.7$ dB, $13$ of $20000$ frames fail, so the FER is $6.5\times10^{-4}$. From $24.8$ dB, no frame fails in $20000$ frames, so the FER is below $1.8\times10^{-4}$ with $95\%$ confidence. All failures start in the 16QAM layer and propagate to the 64QAM layer. At $25$ dB, $\log_2(1+a_2)=6.733$. Since a single layer has $\mathcal D_{\mathcal I_2}(f)=0$, no single-stream scheme, synchronous or asynchronous, with any spectrum and any receiver, can provide a higher rate for User 2. An additional $0.967$ bits are provided to User 2 by the coded link. To give User 2 the rate $7.7$, even when User 1 is silent, a single-stream scheme would need $27.9$ dB, i.e., $3.1$ dB more than the coded link at $24.8$ dB. Fig.~\ref{fig:coded}(a) shows that the offered rate pair lies outside the single-stream regions and inside the A-RS regions. 
 In other words, theoretically, no single-stream scheme can achieve the pair of rates that this practical design achieves. 

\section{Conclusion}
\label{sec:conclusion}

We studied asynchronous rate splitting for the two-user uplink. In the frequency domain, the layered channel is a family of memoryless MACs whose Gram matrices are sums of rank-one alias terms. Therefore, timing offsets create additional signal dimensions only in the folded excess band. Every subset rate has a closed form in terms of a timing dispersion. The layer-rate region is a polymatroid attained by vector SIC. The user-rate region of a fixed design is a pentagon. Two layers per user, separated by half a symbol, maximize all dispersions for any delay between the users. With the proposed nested water-filling, the A-RS region is a pentagon. This region contains the regions of synchronous NOMA and A-NOMA, and for $\beta>0$, its single-user rates exceed $\log_2(1+a_k)$. Practical systems using LDPC codes can achieve rate pairs that are impossible for any single-stream system. 
Extensions to more users, fading channels, and imperfect timing are left for future work.

\appendices
\section{}
\label{proof:folded_spectrum}

Since $R_p\in L^1(\R)$, the Poisson summation~\cite{stein2003} applies. Substituting $r_p((m+\Delta)T)=\int_{\R}R_p(F)e^{j2\pi F(m+\Delta)T}\mathrm{d}F$ into \eqref{eq:G_delta_DTFT} and using $\sum_{m\in\Z}e^{j2\pi m(FT-f)}=\frac1T\sum_{\ell\in\Z}\delta(F-\frac{f+\ell}{T})$ gives \eqref{eq:G_delta_alias_expansion} with the weights \eqref{eq:alias_weight_def}. At $\Delta=0$, the Nyquist property gives $G_0(f)=1$, which is \eqref{eq:alias_weight_partition}. The $(i,j)$-th entry of the right-hand side of \eqref{eq:R_frequency_gram_sum} is $\sum_\ell w_\ell(f)e^{j2\pi(f+\ell)(\tau_j-\tau_i)}=G_{\tau_j-\tau_i}(f)$. The remaining properties hold because \eqref{eq:R_frequency_gram_sum} is a nonnegative combination of rank-one Hermitian matrices, $w_\ell$ is real, and $|G_\Delta(f)|\le\sum_\ell w_\ell(f)=1$ by the triangle inequality.

\section{}
\label{proof:rank}
By \eqref{eq:R_frequency_gram_sum}, $\boldsymbol R_{\mathcal{M}}(f)=\boldsymbol V(f)\boldsymbol W(f)\boldsymbol V^{H}(f)$ with $\boldsymbol V(f)$ the matrix whose columns are $\{\boldsymbol v_{\ell}(f)\}_{\ell\in\mathcal L(f)}$ and $\boldsymbol W(f)=\diag\{w_{\ell}(f)\}_{\ell\in\mathcal L(f)}\succ\boldsymbol0$. Since $\boldsymbol W(f)$ is nonsingular, the range of $\boldsymbol R_{\mathcal{M}}(f)$ is equal to the range of $\boldsymbol V(f)$, which proves the equality in \eqref{eq:R_frequency_rank_bound} and the bound follows because $\boldsymbol V(f)$ has $M$ rows and $|\mathcal L(f)|$ columns. If $\mathcal L(f)=\{\ell_0\}$, then \eqref{eq:alias_weight_partition} forces $w_{\ell_0}(f)=1$ and \eqref{eq:G_delta_alias_expansion} gives $G_{\Delta}(f)=e^{j2\pi(f+\ell_0)\Delta}$, whose modulus is one, while $\boldsymbol R_{\mathcal{M}}(f)=\boldsymbol v_{\ell_0}(f)\boldsymbol v_{\ell_0}^{H}(f)$ has rank one. If all offsets coincide, every $\boldsymbol v_{\ell}(f)$ is a unit-modulus multiple of $\boldsymbol1_M$. Therefore, the sum in \eqref{eq:R_frequency_gram_sum} collapses to $\boldsymbol1_M\boldsymbol1_M^{H}$ by \eqref{eq:alias_weight_partition}.

\section{}
\label{proof:two_alias}
By \eqref{eq:G_delta_alias_expansion}, with only two active terms,
\begin{equation}\label{eq:two_alias_G_factored}
G_{\Delta}(f)=e^{j2\pi(f+\ell_1)\Delta}
\Big[w_1(f)+w_2(f)e^{j2\pi(\ell_2-\ell_1)\Delta}\Big].
\end{equation}
Writing $\phi\triangleq2\pi(\ell_2-\ell_1)\Delta$ and using $w_1(f)+w_2(f)=1$,
\begin{align}
|G_{\Delta}(f)|^2
&=w_1^2(f)+w_2^2(f)+2w_1(f)w_2(f)\cos\phi
\nonumber\\
&=\big(w_1(f)+w_2(f)\big)^2-2w_1(f)w_2(f)(1-\cos\phi)
\nonumber\\
&=1-4w_1(f)w_2(f)\sin^2(\phi/2),
\label{eq:two_alias_G_magnitude}
\end{align}
which is \eqref{eq:two_alias_dimension_gain}. The negative term vanishes if and only if $w_1(f)w_2(f)=0$ or $\sin(\phi/2)=0$, the latter being equivalent to $(\ell_2-\ell_1)\Delta\in\Z$. When $|\ell_2-\ell_1|=1$, the factor $\sin^2(\pi\Delta)$ is independent of $f$ and achieves its maximum over $\Delta\in[0,1)$ at $\Delta=1/2$. Therefore, the maximization is pointwise in $f$ and is achieved by the same $\Delta$ for all $f\in\mathcal E$.


\section{}
\label{proof:closed_form}
Write $\boldsymbol V_{\mathcal A}(f)$ for the $|\mathcal A|\times2$ matrix whose columns are the two active alias steering vectors restricted to $\mathcal A$, and $\boldsymbol W(f)=\diag(w_1(f),w_2(f))$. By Lemma~\ref{lem:folded_spectrum}, $\boldsymbol R_{\mathcal A}(f)=\boldsymbol V_{\mathcal A}(f)\boldsymbol W(f)\boldsymbol V_{\mathcal A}(f)^{H}$, so with $\widetilde{\boldsymbol V}(f)\triangleq\boldsymbol D_{\mathcal A}^{1/2}(f)\boldsymbol V_{\mathcal A}(f)$ and Sylvester's identity~\cite{hornjohnson2013},
\begin{equation}\label{eq:closed_form_sylvester}
\det\left(\boldsymbol I+\boldsymbol D_{\mathcal A}^{1/2}\boldsymbol R_{\mathcal A}
\boldsymbol D_{\mathcal A}^{1/2}\right)
=\det\left(\boldsymbol I_2+\boldsymbol W\widetilde{\boldsymbol V}^{H}
\widetilde{\boldsymbol V}\right).
\end{equation}
With $\ell_1,\ell_2$ the active aliases and $\ell_2-\ell_1=\pm1$, the entries of the $2\times2$ Gram matrix are
\begin{equation}\label{eq:closed_form_gram_entries}
\big[\widetilde{\boldsymbol V}^{H}(f)\widetilde{\boldsymbol V}(f)\big]_{rs}
=\sum_{i\in\mathcal A}\rho_i(f)e^{j2\pi(\ell_r-\ell_s)\tau_i}.
\end{equation}
Therefore, the diagonal entries are equal to $\rho_{\mathcal A}(f)$ and the off-diagonal entries are equal to $\rho_{\mathcal A}(f)z_{\mathcal A}^*(f)$ and $\rho_{\mathcal A}(f)z_{\mathcal A}(f)$, where replacing $e^{j2\pi\tau_i}$ by $e^{-j2\pi\tau_i}$ changes $z_{\mathcal A}$ into its conjugate and leaves $|z_{\mathcal A}|$ unchanged. Expanding the $2\times2$ determinant in \eqref{eq:closed_form_sylvester} and using $w_1+w_2=1$,
\begin{align}
\det\left(\boldsymbol I_2+\boldsymbol W\widetilde{\boldsymbol V}^{H}
\widetilde{\boldsymbol V}\right)
&=(1+w_1\rho_{\mathcal A})(1+w_2\rho_{\mathcal A})
-w_1w_2\rho_{\mathcal A}^2|z_{\mathcal A}|^2
\nonumber\\
&=1+\rho_{\mathcal A}+w_1w_2\rho_{\mathcal A}^2\big(1-|z_{\mathcal A}|^2\big),
\label{eq:closed_form_expand}
\end{align}
which is \eqref{eq:Jf_closed_form}. The bound $|z_{\mathcal A}|\le1$ follows from the triangle inequality, giving $\mathcal D_{\mathcal A}\in[0,1]$, with $|z_{\mathcal A}|=1$ if and only if all offsets $\{\tau_i\}_{i\in\mathcal A}$ carrying positive power are equal. Outside $\mathcal E$, one of the weights vanishes and the last term disappears.

\section{}
\label{proof:decomposition}

Conditioning on $\{\boldsymbol s_j\}_{j\notin\mathcal A}$ removes the layers from \eqref{eq:stack} and leaves a channel of the same form with $\mathcal M$ replaced by $\mathcal A$, so it suffices to take $\mathcal A=\mathcal M$. We follow the approach of~\cite{verdu1989asynchronous} for one stream per user and only indicate the changes. In the time-stacked order of Section~\ref{sec:decomposition}, the Gram matrix is the block-Toeplitz matrix $\boldsymbol T_N$ generated by $\boldsymbol R_{\mathcal M}(f)$. The noise has covariance $\sigma^2\boldsymbol T_N$. Whitening it on the range of $\boldsymbol T_N$ and applying Sylvester's identity give, for independent unit-power Gaussian inputs~\cite{telatar1999capacity},
\begin{equation}\label{eq:IN_block_toeplitz}
I_N=\log_2\det\left(\boldsymbol I_{MN}+\boldsymbol D_N^{1/2}\boldsymbol T_N\boldsymbol D_N^{1/2}\right),
\end{equation}
where $\boldsymbol D_N=\boldsymbol I_N\otimes\diag(\gamma_1,\ldots,\gamma_M)$, since the phases of $g_i$ do not change the determinant. The matrix in \eqref{eq:IN_block_toeplitz} is Hermitian block Toeplitz, generated by the continuous matrix $\boldsymbol I_M+\boldsymbol D^{1/2}\boldsymbol R_{\mathcal M}(f)\boldsymbol D^{1/2}$, $\boldsymbol D=\diag(\gamma_1,\ldots,\gamma_M)$. Its eigenvalues lie in $[1,1+M\max_i\gamma_i]$, since $\boldsymbol T_N\succeq\boldsymbol 0$ and $\|\boldsymbol R_{\mathcal M}(f)\|_2\le\tr\boldsymbol R_{\mathcal M}(f)=M$. The block Szeg\H{o} theorem~\cite{gutierrez2008toeplitz} applied to the logarithm, which is continuous on this interval, results in
\begin{align}
\lim_{N\to\infty}\frac{I_N}{N}&=\int_{\mathcal F}\log_2\det\left(\boldsymbol I_M+\boldsymbol D^{1/2}\boldsymbol R_{\mathcal M}(f)\boldsymbol D^{1/2}\right)\mathrm{d}f
\nonumber\\
&=J(\mathcal M).
\end{align}

\section{}
\label{proof:shaped_decomposition}
As before, it suffices to take $\mathcal A=\mathcal M$. With the DFT precoding of Algorithm~\ref{alg:vector_sic}, the input covariance is block circulant, and the block DFT diagonalizes it into $\diag(Q_1(f_n),\ldots,Q_M(f_n))$, $f_n=n/N$. As in the proof of Theorem~\ref{thm:decomposition}, $I_N$ is the log-determinant of the identity plus a matrix that involves $\boldsymbol T_N$ and this covariance. We replace $\boldsymbol T_N$ with the block-circulant matrix $\boldsymbol C_N$ generated by the periodized blocks $\sum_{z\in\Z}\boldsymbol R[r+zN]$. Absolute summability gives $\|\boldsymbol T_N-\boldsymbol C_N\|_F^2/N\to0$~\cite{gray2006toeplitz}. Since all matrices involved have uniformly bounded norms and the eigenvalues of both log-determinant arguments are at least one, the Hoffman-Wielandt inequality~\cite{hoffman1953spectrum} shows that the two normalized log-determinants differ by $o(1)$. The block DFT diagonalizes $\boldsymbol C_N$ into blocks $\boldsymbol R_{\mathcal M}(f_n)$. Thus, the circulant log-determinant is $\sum_nJ_{f_n}(\mathcal M)$, evaluated with $\boldsymbol D(f_n)=\diag(\rho_1(f_n),\ldots,\rho_M(f_n))$. Since each $\rho_i$ is bounded and Riemann integrable and $\boldsymbol R_{\mathcal M}(f)$ is continuous, $J_f(\mathcal M)$ is Riemann integrable~\cite{rudin1976analysis}, and the Riemann sum $\frac1N\sum_nJ_{f_n}(\mathcal M)$ converges to $J(\mathcal M)$.

\section{}
\label{proof:polymatroid}

It suffices to prove the statement for $J_f$, since integration over $\mathcal F$ preserves all three properties. With $\boldsymbol V(f)$ and $\boldsymbol W(f)$ as in the proof of Lemma~\ref{lem:rank}, consider the auxiliary memoryless channel
\begin{equation}\label{eq:aux_channel_short}
\boldsymbol y_f=\sum_{i\in\mathcal M}\boldsymbol b_i(f)s_i+\boldsymbol\nu_f,
\end{equation}
with $\boldsymbol b_i(f)=\frac{g_i}{|g_i|}\sqrt{\rho_i(f)}\boldsymbol W^{1/2}(f)\boldsymbol V^H(f)\boldsymbol e_i$ of Algorithm~\ref{alg:vector_sic}, independent unit-variance proper Gaussian inputs $s_i$ and standard proper Gaussian noise $\boldsymbol\nu_f$. Eq. \eqref{eq:aux_channel_short} is the observation of Algorithm~\ref{alg:vector_sic} in one frequency bin. Since the vectors $\boldsymbol W^{1/2}\boldsymbol V^H\boldsymbol e_i$ have Gram matrix $\boldsymbol R_{\mathcal M}(f)$, Sylvester's identity gives $I(s_{\mathcal A};\boldsymbol y_f\mid s_{\mathcal A^c})=J_f(\mathcal A)$. The conditional mutual informations of a MAC with independent inputs form a polymatroid rank function~\cite{tsehanly1998polymatroid}. Normalization is immediate, monotonicity follows from the chain rule, and submodularity follows from $I(U;V\mid\boldsymbol y_f,W)\ge0=I(U;V\mid W)$ for disjoint groups $U$, $V$, $W$ of independent inputs.

\section{}
\label{proof:layer_polymatroid}

For the converse, given the layers in $\mathcal A^c$, the layers in $\mathcal A$ cannot carry more than $I(\boldsymbol s_{\mathcal A};\boldsymbol y\mid\boldsymbol s_{\mathcal A^c})$~\cite{cover2006elements}, whose normalized limit is $J(\mathcal A)$ by Theorem~\ref{thm:decomposition} or Corollary~\ref{cor:shaped_decomposition}. For achievability, at Stage $m$ of SIC, the layers in $\mathcal S_m^c$ have been canceled, and those in $\mathcal S_{m+1}$ act as Gaussian noise. By the chain rule, $I(\boldsymbol s_{\pi(m)};\boldsymbol y\mid\boldsymbol s_{\mathcal S_m^c})=I(\boldsymbol s_{\mathcal S_m};\boldsymbol y\mid\boldsymbol s_{\mathcal S_m^c})-I(\boldsymbol s_{\mathcal S_{m+1}};\boldsymbol y\mid\boldsymbol s_{\mathcal S_{m+1}^c})$, whose normalized limit is \eqref{eq:exact_sic_rate}. These rates form the greedy vector of $J$ for the reversed order $\pi(M),\ldots,\pi(1)$. Therefore, by Lemma~\ref{lem:polymatroid} and Edmonds' theorem~\cite{edmonds1970submodular}, they are vertices of the dominant face of $\mathcal P(J)$, and every such vertex arises from some $\pi$. Time sharing and rate reduction attain the rest of $\mathcal P(J)$. Summing \eqref{eq:exact_sic_rate} over $m$ telescopes to \eqref{eq:sic_telescoping}.

\section{}
\label{proof:aggregation}

The inclusion $\subseteq$ follows from the constraints of $\mathcal P(J)$ for $\mathcal A=\mathcal I_1$, $\mathcal I_2$, and $\mathcal M$. Conversely, monotonicity and submodularity, $J(\mathcal M)\le J(\mathcal I_1)+J(\mathcal I_2)$, show that the right-hand side is the pentagon with corners $\boldsymbol\zeta_1=\big(J(\mathcal I_1),J(\mathcal M)-J(\mathcal I_1)\big)$ and $\boldsymbol\zeta_2=\big(J(\mathcal M)-J(\mathcal I_2),J(\mathcal I_2)\big)$. The greedy vector of an order that lists the layers of $\mathcal I_1$ first has the partial sums $J(\mathcal I_1)$ over $\mathcal I_1$ and $J(\mathcal M)$ over $\mathcal M$, so it projects onto $\boldsymbol\zeta_1$. By the proof of Theorem~\ref{thm:layer_polymatroid}, it is attained by every SIC order that decodes all layers of $\mathcal I_2$ before those of $\mathcal I_1$. Exchanging the users gives $\boldsymbol\zeta_2$. The left-hand side is convex and closed under rate reduction, so it contains the pentagon.

\section{}
\label{proof:greedy}
Consider $\mu_1\ge\mu_2$ (the case $\mu_2\ge\mu_1$ follows by exchanging the users). By Lemma~\ref{lem:aggregation}, every $(R_1,R_2)\in\mathcal C$ satisfies $R_1\le J(\mathcal I_1)$ and $R_1+R_2\le J(\mathcal M)$. Since $\mu_2\ge0$ and $\mu_1-\mu_2\ge0$, it follows that
\begin{align}
	\mu_1R_1+\mu_2R_2
	&=
	\mu_2(R_1+R_2)+(\mu_1-\mu_2)R_1
	\nonumber\\
	&\le
	\mu_2J(\mathcal M)+(\mu_1-\mu_2)J(\mathcal I_1).
	\label{eq:weighted_objective_derivation}
\end{align}
The bound is achieved at the vertex $\boldsymbol\zeta_1=\big(J(\mathcal I_1),J(\mathcal M)-J(\mathcal I_1)\big)$ of $\mathcal C$, where both constraints are  held with equality. This proves the first line of \eqref{eq:weighted_objective}.

By the proof of Lemma~\ref{lem:aggregation}, $\boldsymbol\zeta_1$ is attained by every SIC order that decodes all layers of $\mathcal I_2$ before those of $\mathcal I_1$. User~1 is then decoded last, after all layers of User~2 have been canceled. User 1's layers receive the total $J(\mathcal I_1)$, while those of User~2 receive the remainder $J(\mathcal M)-J(\mathcal I_1)$. These group totals do not depend on the order within each group, and therefore neither does $h_{\mathcal C}(\boldsymbol\mu)$. Thus, the higher-weight user is decoded last, where it benefits from the cancellation of all other layers.

\section{}
\label{proof:diagonal_solution}

For each $f$ and $x\ge0$, let $g_f(x)\triangleq\frac{1+2w(f)x}{1+x+w(f)x^2}$, which is $\ln2$ times the derivative of the integrand of \eqref{eq:V_def} at $\rho(f)=x$. Since
\begin{equation}
g_f'(x)=-\frac{1-2w(f)+2w(f)x+2w^2(f)x^2}{\big(1+x+w(f)x^2\big)^2}<0,
\end{equation}
for $w(f)\le1/4$, the integrand is strictly concave. Thus, the KKT conditions are necessary and sufficient and the maximizer is unique. The Lagrangian separates over $f$, and with $\nu$ equal to $\ln2$ times the multiplier, the conditions are $g_f(\rho(f))=\nu$ if $\rho(f)>0$ and $g_f(0)=1\le\nu$ if $\rho(f)=0$. For $a>0$, we need $\nu<1$, since $g_f(x)<1$ for $x>0$. Then, $\rho(f)>0$ at every $f$, and $g_f(\rho)=\nu$ becomes $\nu w(f)\rho^2+(\nu-2w(f))\rho+\nu-1=0$, whose unique positive root is \eqref{eq:diagonal_kkt}. For $w(f)=0$, the root is $1/\nu-1$. The power used is continuous and strictly decreasing in $\nu$, from infinity as $\nu\to0$ to zero as $\nu\to1$, so exactly one $\nu$ meets the budget.

\section{}
\label{proof:excess_band_doubling}

Let $\nu_a$ be the multiplier of Proposition~\ref{prop:diagonal_solution} for the budget $a$. Since $1/(1+x)\le g_f(x)\le2/x$ for $x>0$, we have $1/\nu_a-1\le\rho_a^\star(f)\le2/\nu_a$. Thus, $\nu_a\to0$ and $\rho_a^\star(f)\to\infty$ as $a\to\infty$. Then,
\begin{equation}
\nu_a\rho_a^\star(f)=\frac{\rho_a^\star(f)+2w(f)\rho_a^{\star2}(f)}{1+\rho_a^\star(f)+w(f)\rho_a^{\star2}(f)}
\end{equation}
tends to $2$ on $\mathcal E$ and to $1$ outside it, and since $0\le\nu_a\rho_a^\star(f)\le2$, dominated convergence gives $a\nu_a\to1+\beta$, which proves \eqref{eq:doubling}. With $r_a(f)\triangleq(1+\beta)\rho_a^\star(f)/a$, the integrand of $V(a)$ is equal to $\log_2w(f)+2\log_2\frac{a}{1+\beta}+2\log_2r_a(f)+o(1)$ in $\mathcal E$ and $\log_2\frac{a}{1+\beta}+\log_2r_a(f)+o(1)$ outside it. Since $\log_2r_a(f)$ tends to $1$ in $\mathcal E$ and to $0$ outside it, $|\mathcal E|=\beta$, and $\log_2w$ is integrable on $\mathcal E$,
\begin{equation}
V(a)=(1+\beta)\log_2\frac{a}{1+\beta}+2\beta+\int_{\mathcal E}\log_2w(f)\mathrm{d}f+o(1).
\end{equation}
For the RRC pulse, each folded edge interval of $\mathcal E$, parameterized by $u\in[0,1]$, has $w(f)=\frac14\sin^2(\pi u/2)$. Thus, $\int_{\mathcal E}\log_2w(f)\mathrm{d}f=\beta\int_0^1\log_2\big(\frac14\sin^2\frac{\pi u}{2}\big)\mathrm{d}u=-4\beta$~\cite{gradshteyn2014table}, which gives \eqref{eq:V_asymptote}.

\bibliography{journalAbbreviations, references}
\bibliographystyle{ieeetr}

\end{document}